\documentclass[useAMS,usenatbib]{mn2e}
 \usepackage{graphicx}
 \usepackage{amsmath}

\def\OI{[\mbox{O\,{\sc i}}]~$\lambda$6300}
\def\OII{[\mbox{O\,{\sc ii}}]}

\def\OIIIHb{[\mbox{O\,{\sc iii}}]/H$\beta$}
\def\OIII{[\mbox{O\,{\sc iii}}]}
\def\OIII5007Hb{[\mbox{O\,{\sc iii}}]~$\lambda5007$/H$\beta$}
\def\OIII49595007{[\mbox{O\,{\sc iii}}]~$\lambda \lambda 4959,5007$}
\def\Zsun{${\rm Z}_{\odot}$}
\def\Msun{${\rm M}_{\odot}$}
\def\Lsun{${\rm L}_{\odot}$}
\def\NII{[\mbox{N\,{\sc ii}}]}
\def\lOII{[\mbox{O\,{\sc ii}}]~$\lambda \lambda$3726,9}
\def\OIIIOII{[\mbox{O\,{\sc iii}}]/[\mbox{O\,{\sc ii}}]}

\def\NIIHa{[\mbox{N\,{\sc ii}}]/H$\alpha$}
\def\SIIHa{[\mbox{S\,{\sc ii}}]/H$\alpha$}
\def\SIIl{[\mbox{S\,{\sc ii}}]~$\lambda \lambda 6717,31$}
\def\OIHa{[\mbox{O\,{\sc i}}]/H$\alpha$}
\def\SII{[\mbox{S\,{\sc ii}}]}
\def\Hb{{H$\beta$}}
\def\Hd{{H$\delta$}}
\def\HdA{{H$\delta_{A}$}}
\def\Ha{{H$\alpha$}}

\def\Lsun{L$_{\odot}$}

\def\D4{$D4000$}

\def\LOIII{L[\mbox{O\,{\sc iii}}]}
\def\LOIIIsig{L[\mbox{O\,{\sc iii}}]/$\sigma^4$}
\def\SFD{${\rm D}_{SF}$}
\def\DSF{${\rm D}_{SF}$}
\def\vel{$\sigma_{*}$}

\title[Host Galaxies of Active Galactic Nuclei]{The Host Galaxies and Classification of Active Galactic Nuclei}

\author[L. Kewley et al.]{Lisa J. Kewley$^{1}$\thanks{Hubble Fellow; kewley@ifa.hawaii.edu}\\
University of Hawaii, 2680 Woodlawn Drive, Honolulu, HI 96822, USA
\newauthor
Brent Groves,  Guinevere Kauffmann \\ Max Plank Institut fur Astrophysik
\newauthor
Tim Heckman\\
Johns Hopkins University}

\begin{document}

\date{Submitted 2006 May 26}

\maketitle
\begin{abstract}
We present an analysis of the host properties of 85224 emission-line
galaxies selected from the Sloan Digital Sky Survey. We show that Seyferts
and LINERs form clearly separated branches on the standard optical
diagnostic diagrams. We derive a new empirical classification scheme
which cleanly separates star-forming galaxies, composite AGN-\mbox{H\,{\sc ii}} galaxies, 
Seyferts and LINERs and we study the host galaxy properties of these
different classes of objects. LINERs are older, more massive, less dusty and
more concentrated, and they and have higher velocity dispersions
and lower [OIII] luminosities than Seyfert galaxies. 
Seyferts and LINERs are most strongly distinguished by their [OIII] luminosities.
We then consider the  quantity L[OIII]/$\sigma^4$, which is an  
indicator of the black hole accretion rate relative to the Eddington rate.
Remarkably, we find that at {\it fixed} L[OIII]/$\sigma^4$,
all differences between Seyfert and LINER  host properties
disappear. LINERs and Seyferts form a continuous sequence,
with LINERs dominant at low $L/L_{EDD}$  and Seyferts
dominant at high $L/L_{EDD}$ . These results suggest that the majority of LINERs are 
AGN and that the Seyfert/LINER dichotomy is analogous to the high/low-state
transition for X-ray binary systems.  We apply theoretical photo-ionization
models and show that pure LINERs require a harder ionizing
radiation field with lower ionization parameter than Seyfert galaxies, consistent
with the low and high X-ray binary states.
\end{abstract}

\nokeywords

\section{Introduction}

The majority of nearby AGN and AGN candidates have nuclear optical spectra that are dominated by emission lines of low ionization species such as [OI]~$\lambda$6300, \lOII\ and \SIIl \citep{Ho97b}.  This class of AGN 
was first defined by \citet{Heckman80} as LINERs (Low Ionization Narrow Emission-line Regions).  LINERs have lower luminosities than Seyfert galaxies or quasars and are therefore often referred to as low-luminosity active galactic nuclei (LLAGN).    LINER emission is extremely common in the nuclei of galaxies; up to $\sim 1/3$ of all galaxies have nuclear spectra typical of LINERs \citep{Heckman80,Ho95, Ho97}.  Despite the prevalence of LINERs in galaxies and decades of study, the power source of LINERs is still under debate.  Some LINER galaxies have double-peaked broad Balmer lines \citep{Eracleous01,Storchi97,Bower96}, while others have compact radio cores \citep{Falcke00,Ulvestad01,Filho02,Filho04,Anderson04}, evidence for hard X-ray spectra \citep{Terashima00,Ho01}, and/or UV variability \citep{Maoz05}.  These observations provide circumstantial evidence for an AGN power source for the LINER emission.  Other possible power 
sources include fast shocks \citep{Heckman80,Dopita95,Lipari04}, photoionization by hot stars \citep{Filippenko92,Shields92,Maoz98,Barth00}, or photoionization by an old, metal-rich stellar population \citep{Taniguchi00,Alonso00}.  LINER emission has been observed in extranuclear regions associated with large-scale outflows and related shocks \citep{Lipari04}, or regions shocked by radio jets \citep{Cecil00}.

Recent investigations into the stellar populations of LINER host galaxies have yielded important insight into their nature. 
 \citet{Maoz98} and \citet{Colina02} detected stellar wind lines in UV spectra of
weak \OIHa\ LINERs.  Similar features have been 
detected in some Seyfert galaxies from nuclear starbursts that are a few Myr old \citep{Heckman97,Gonzalez98}.
\citet{Gonzalez04} searched the HST STIS spectra of 28 LINERs for Wolf-Rayet features.  They found no Wolf-Rayet features and little evidence of young stars in LINERs with 
strong \OIHa.  In LINERs with low \OIHa\ intermediate-age stars contribute significantly to 
the stellar continuum.  
\citet{Fernandes04} found  that while massive stars do not contribute significantly to LINER 
spectra, high order Balmer absorption lines are detected in $\sim 50$\% of LINERs that have relatively weak \OI\ emission.  These results indicate that the current LINER classification scheme encompasses two or more types of galaxies, or galaxies at different stages in evolution.  

LINERs are commonly classified using their optical emission-line ratios.  The first optical classification classification scheme to segregate LINERs from other spectral types was proposed by \citet{Heckman80}.  This scheme uses line ratios of the low ionization species  [\mbox{O\,{\sc i}}] $\lambda 6300$ and \OII~$\lambda \lambda 3726,29$ compared to the high ionization species  [\mbox{O\,{\sc iii}}] $\lambda 5007$ to separate 
LINERs from Seyferts.   This scheme requires the use of additional diagnostics to remove star-forming galaxies.  The  most common method to remove star-forming galaxies is based on the \citet{Baldwin81} empirical diagnostic diagrams using the optical line ratios \OIHa, \SIIHa, \NIIHa, and \OIIIHb.   The Baldwin et al. scheme was revised by \citet{Osterbrock85} and \citet{Veilleux87}.   An alternative scheme was proposed by \citet{Ho97} that includes an additional
division between ``pure'' LINERs and LINER/\mbox{H\,{\sc ii}}\ ``transition'' objects using the \OIHa\ ratio.  Transition objects have line ratios that are intermediate between the two classes.  This division is arbitrary because there was no clear division in \OIHa\ between transition or true LINER classes.  The first purely theoretical classification scheme to distinguish between AGN, LINERs, and \mbox{H\,{\sc ii}} region-like 
objects was derived by \citet{Kewley01b}.  They used a combination of modern stellar population synthesis, photoionization, and shock models to derive a "maximum starburst line" and an "extreme mixing line" for separation of the three types of objects.   Kewley et al. concluded that previous LINER classification schemes include starburst-Seyfert composites in the LINERs class, as well as bona-fide LINERs.   Recently, \citet{Kauffmann03b} shifted the Kewley et al. extreme starburst line to make a semi-empirical fit to the outer bound of $\sim 22,600$ SDSS spectra.  This outer bound defines the region where composite starburst-AGN objects are expected to lie on the diagnostic diagrams.

Although much progress has been made in the optical classification of the ionizing source in galaxies, none of these classification schemes has been able to divide cleanly between Seyfert, LINER and composite/transition types.   The primary reason for this problem is the lack of a sufficiently large sample in which empirical boundaries between the different galaxy classes can be observed.  In this paper, we use 85,224 galaxies from the Sloan Digital Sky Survey (SDSS) data release 4 (DR4) emission-line catalog (described in Section~\ref{sample}) to develop a new semi-empirical classification scheme for Seyferts, LINERs, and composite objects (Section~\ref{class}).  We use this new classification scheme to investigate the host properties of AGN in Section~\ref{host}.  Our results are discussed in 
Section~\ref{discussion} and we present our conclusions in Section~\ref{conclusions}.

Throughout this paper, we adopt the flat 
$\Lambda$-dominated cosmology as measured by the WMAP experiment 
($h=0.72$, $\Omega_{m}=0.29$; \citealt{Spergel03}).

\section{Sample Selection}\label{sample}

Our sample was selected from the 567486-galaxy DR4 sample according to the 
following criteria:

\begin{enumerate}
\item Signal-to-noise ratio S/N$\ge3$ in the strong emission-lines \Hb, \mbox{O\,{\sc iii}}~$\lambda 5007$, 
\Ha, \NII~$\lambda 6584$, and \SII~$\lambda \lambda 6717,31$.  
\item Redshifts between $0.04<z<0.1$.
\end{enumerate}

The S/N criterion is required for accurate classification of the galaxies into star forming or AGN dominated classes (e.g., \citealt{Kewley01a,Veilleux87}).   Our lower redshift limit ensures that the galaxy properties derived from the fiber spectra are not dominated by the small fixed-size aperture.
\citet{Kewley05a} analysed the effect of a fixed size 
aperture on metallicity, star-formation rate, and reddening.  They concluded that a minimum aperture covering fraction of $\sim 20$\% is required for the spectral properties within the aperture to approximate 
the global values.   For the $3^{\prime \prime}$ fiber aperture of the SDSS,  a 20\% covering fraction 
corresponds roughly to a redshift of $z\sim0.04$.  Our upper redshift limit avoids incompleteness in the LINER class.  LINERs typically have lower luminosities than Seyfert galaxies, and are therefore found at lower redshifts than Seyferts in the magnitude-limited SDSS survey.  We will investigate incompleteness as a function of galaxy type in Section~\ref{new_class}

The resulting sample contains 85224 emission-line galaxies and does not include duplicates found in the original DR4 catalog.  We use the publically available emission-line fluxes (described in \citealt{Tremonti04}).  These fluxes were calculated using a
sophisticated technique that applies a least-squares fit of stellar population synthesis models and dust attenuation to the continuum.  Once the continuum has been removed, the emission-line fluxes were fit with Gaussians, constraining the width and velocity separation of the Balmer lines together, and similarly for the forbidden lines.  

We have corrected the emission-line fluxes for extinction using the Balmer decrement and the \citet{Cardelli89} reddening curve.  We assume an ${\rm R_{V}=Av/{\rm E}(B-V)} = 3.1$ and an 
intrinsic H$\alpha$/H$\beta$ ratio of 2.85  for galaxies dominated by star formation and H$\alpha/$H$\beta = 3.1$ for galaxies dominated by AGN \citep[the Balmer decrement for case B 
recombination at T$=10^4$K and $n_{e} \sim 10^2 - 10^4 {\rm cm}^{-3}$;][]{Osterbrock89}.  A total of 5414 (6\%) of galaxies in our sample have Balmer decrements less than the theoretical value.  A Balmer decrement
less than the theoretical value can result
from one or a combination of (1) intrinsically low reddening, (2) errors in the
stellar absorption correction, and (3) errors in the line flux calibration and 
 measurement.  For the S/N of our data, the lowest E(B-V) measurable is 0.01.  We therefore
assign these 5414 galaxies an upper limit of E(B-V)$<0.01$.  

\begin{figure*}
\begin{minipage}{155mm}
\includegraphics[width=15cm]{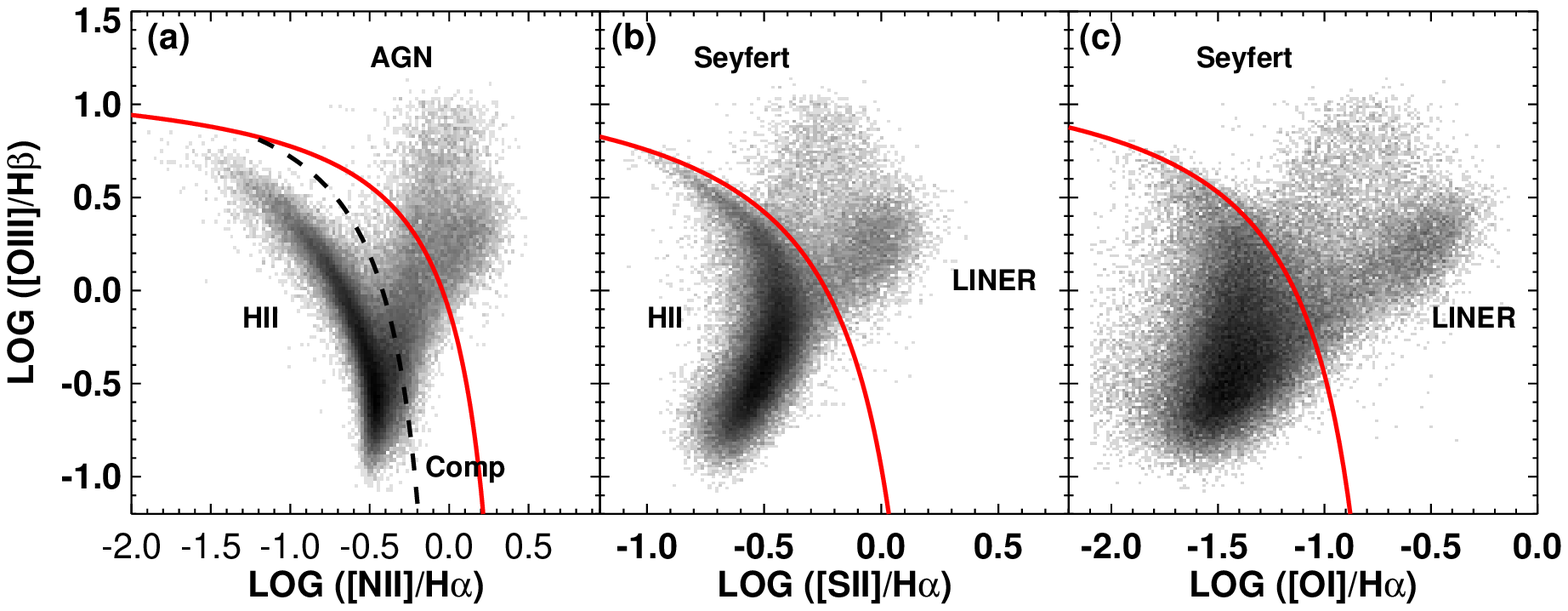}
\caption{(a) The \NIIHa\ vs \OIIIHb\ diagnostic diagram for SDSS galaxies with 
S/N $> 3$.   The \citet{Kewley01a} extreme starburst line and the \citet{Kauffmann03b} classification line are shown as solid and dashed lines respectively. (b) The \SIIHa\ vs \OIIIHb\ diagnostic diagram, 
(c) The \OIHa\ vs \OIIIHb\ diagnostic diagram.
\label{BPT}}
\end{minipage}
\end{figure*}

\section{Optical Classification}\label{class}

\citet{Baldwin81} proposed a suite of three diagnostic diagrams to classify 
the dominant energy source in emission-line galaxies.  These diagrams are 
commonly known as BPT diagrams and are based on the four optical line 
ratios \OIIIHb,\NIIHa, \SIIHa, and \OIHa.    \citet[][; hereafter Ke01]{Kewley01a} used a combination of stellar population synthesis models and detailed self-consistent photoionization models to create a theoretical ``maximum starburst line'' on the the BPT diagrams.   The maximum starburst  line is determined by the upper limit of the theoretical pure stellar photoionization models.  Galaxies lying above this line are likely to be dominated by an AGN.   To rule out possible composite galaxies, \citet[][; hereafter Ka03]{Kauffmann03b} modified the Ke01 scheme to include an empirical line dividing pure star-forming galaxies from Seyfert-\mbox{H\,{\sc ii}} composite objects whose spectra contain significant contributions from both AGN and star formation.

Figure~\ref{BPT}a shows the \OIIIHb\ versus \NIIHa\ standard optical diagnostic
diagram for our sample.  The Ke01 and Ka03 classification lines are shown as solid and dashed lines.  Galaxies that lie below the dashed Ka03 line are classed as \mbox{H\,{\sc ii}} region-like galaxies.  Star-forming galaxies form a tight sequence from low metallicities (low \NIIHa, high \OIIIHb) to high metallicities (high \NIIHa, low \OIIIHb) which we will refer to as the "star-forming sequence".
The AGN mixing sequence begins at the high metallicity end of the star-forming sequence and 
extends towards high \OIIIHb\ and \NIIHa\ values.  Galaxies that lie in between the two classification lines are on the 
AGN-\mbox{H\,{\sc ii}} mixing sequence and are classed as composites.  Composite galaxies are likely to contain a metal-rich stellar population plus an AGN.  Galaxies that lie above the Ka03 line are classed as AGN.

Figures~\ref{BPT}b and ~\ref{BPT}c show the \OIIIHb\ versus \SIIHa\ and 
\OIIIHb\ versus \OIHa\ diagrams for our sample.  The Ke01 classification line provides an 
upper bound to the star-forming sequence on these diagrams.    Galaxies that are classed 
as composites using the \NIIHa\ diagram (Figure~\ref{BPT}a) lie mostly within the 
star-forming sequence on the \SIIHa\ diagram (Figure~\ref{BPT}b) and about half-way into the star-forming sequence in the \OIHa\ (Figure~\ref{BPT}c) diagram.  
The \NIIHa\ line ratio is more sensitive to the presence of low-level AGN than 
\SIIHa\ or \OIHa\ thanks primarily to the metallicity sensitivity of \NIIHa.  The log(\NIIHa) line ratio is a linear function of the nebular metallicity until high metallicities where the log(\NIIHa) ratio saturates \citep{Kewley02a,Denicolo02,Pettini04}.  This saturation point causes the star-forming sequence to be almost vertical at  \NIIHa$\sim -0.5$.  Any AGN contribution shifts
the \NIIHa\ towards higher values than this saturation level, allowing removal of galaxies with even small AGN contributions.

In this work, we classify pure star-forming galaxies as those that lie below the Ka03 line on the \NIIHa\ vs \OIIIHb\ diagnostic diagram.   Composite galaxies lie above the Ka03 line and below the 
Ke01 line.   The optical spectra of composites 
can be due to either (1) a combination of star-formation and  a Seyfert nucleus, or (2) a combination of star-formation and LINER emission.  The narrow line emission from galaxies lying above the Ke01 line is likely to be dominated by an AGN.

\subsection{New LINER, Seyfert, Composite Classification Scheme}\label{new_class}

\begin{figure}
\includegraphics[width=8.5cm]{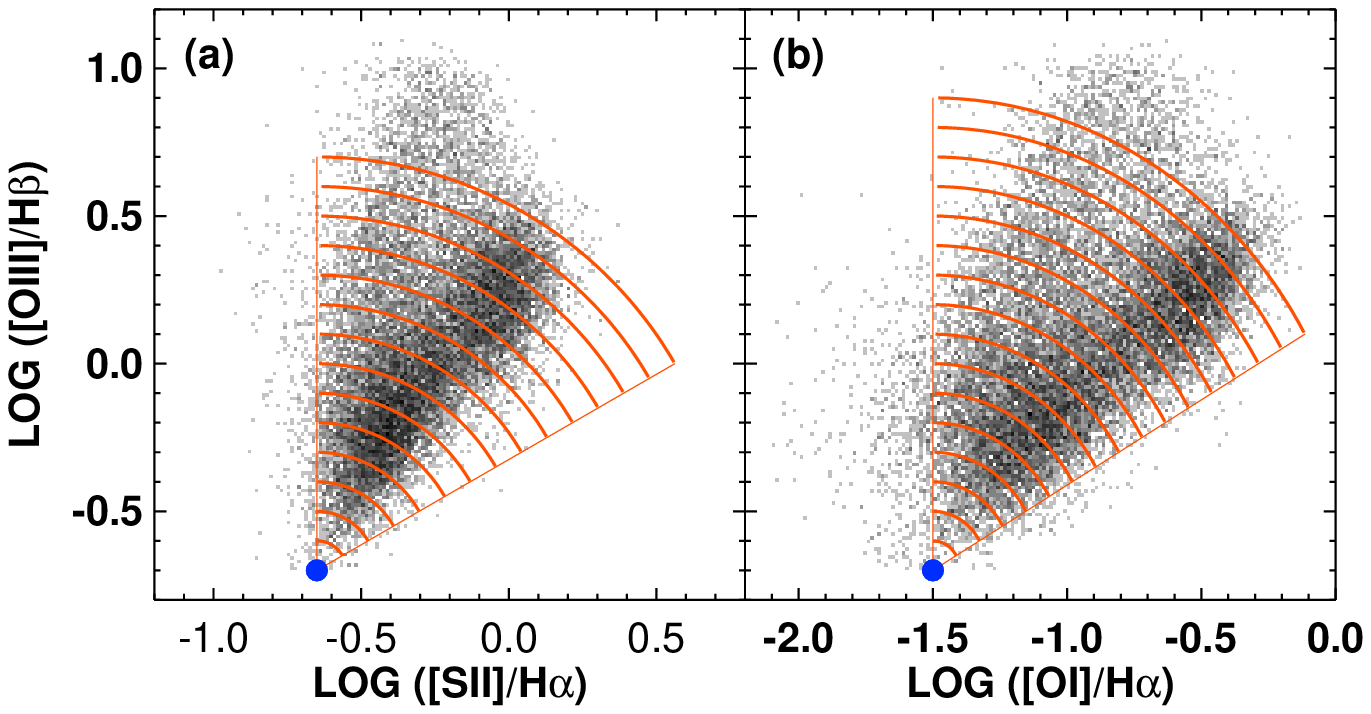}
\caption{(a) The \SIIHa\ vs \OIIIHb\ and (b) \OIHa\ vs \OIIIHb\ (right) diagnostic diagrams 
for SDSS galaxies classified as AGN using the \citet{Kauffmann03b} line (dashed line in 
Figure~\ref{BPT}a.  The large filled circle represents the empirical base point $p$ 
for the Seyfert and LINER sequences.  Concentric arcs of 0.1 dex (red solid lines) show the binning 
of our sample with radius.  We calculate histograms of angle from the x-axis centered at $p$ for each bin.
\label{arcs}}
\end{figure}

\begin{figure}
\includegraphics[width=8.5cm]{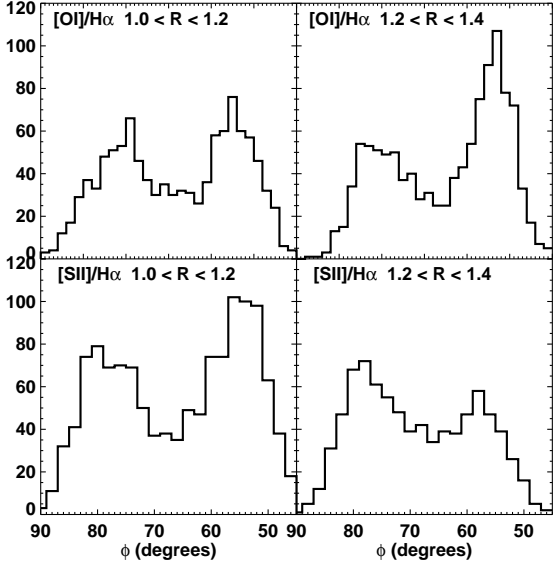}
\caption{Histograms of the Seyfert and LINER sequences between 1.0-1.2 dex (left) and 1.2-1.4 dex (right) for the \OIHa\ (top) and \SIIHa\ (bottom) diagnostic diagrams.  The distribution is clearly bimodal.  The LINER sequence is the right-hand peak and the Seyfert sequence is the left-hand peak.  \label{angle_hist}}
\end{figure}

It is clear from the \SIIHa\ and \OIHa\ diagrams (Figures~\ref{BPT}b and \ref{BPT}c) that galaxies containing AGN lie on two branches.  Seyfert galaxies lie on the upper branch while LINERs lie on 
the lower branch, thanks to their low ionization line emission.   We use the \SIIHa\ and 
\OIHa\ diagrams to separate Seyfert from LINER galaxies.  We use only galaxies with S/N$>6$ in each of the strong lines to derive our new classification scheme.  We define an empirical base point, $p$ (blue circles in Figure~\ref{arcs}) and we define annuli with widths of 0.1 dex centered on $p$.  For the galaxies in each of these annuli with radii (R) for which the two AGN branches are well-defined (0.5-1.5 dex from $p$),  we compute histograms of angle from the x-axis.   We then step the 0.1 dex annuli through the data from $p$ in 0.02 dex increments and recalculate the minima of each histogram for each increment.   A total of 84 histograms were created and minima between the Seyfert and LINER curves were found for all histograms.   Examples of these histograms are given in Figure~\ref{angle_hist}.

\begin{figure*}
\begin{minipage}{155mm}
\includegraphics[width=15cm]{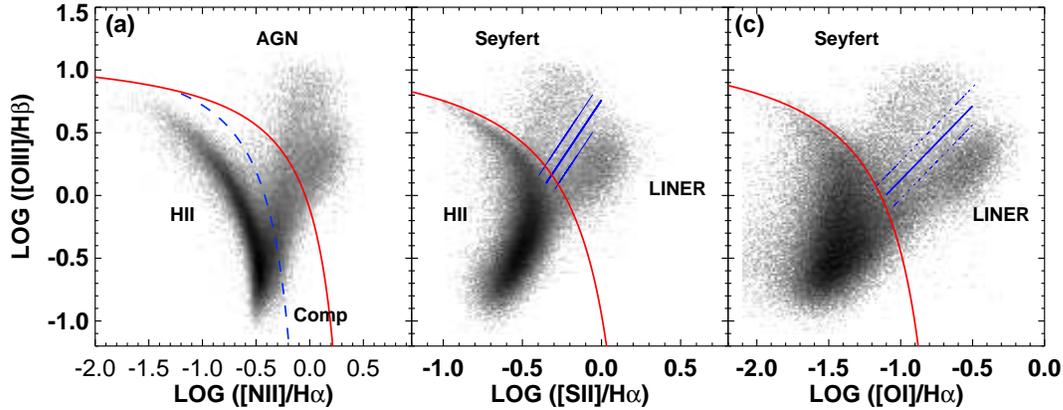}
\caption{The three BPT diagrams showing our new scheme for classifying galaxies using 
emission-line ratios.  The \citet{Kewley01a} extreme starburst classification line (red solid), 
the \citet{Kauffmann03b} pure star-formation line (blue dashed), and our new 
Seyfert-LINER line (blue solid) are used to separate galaxies into \mbox{H\,{\sc ii}} region-like, Seyferts,
LINERs, and Composite \mbox{H\,{\sc ii}}-AGN types.
\label{class_diag}}
\end{minipage}
\end{figure*}

To divide between the two AGN branches, we take the line of best-fit through the histogram minima for each of the two diagnostic diagrams.  This line provides an empirical division between Seyferts and LINERs and will be referred to as the "Seyfert-LINER classification line".  Adding this line to our previous two classification lines from Ke01 and Ka03, we obtain a means to distinguish between 
star-forming galaxies, Seyferts, LINERs, and composite galaxies. Our new classification scheme is shown in Figure~\ref{class_diag}, and 
is defined as follows:

1. {\bf Star-forming galaxies} lie below and to the left of the Ka03 classification line in the \NIIHa\ vs \OIIIHb\ diagram and below and to the left of the Ke01 line in the \SIIHa\ and \OIHa\ diagrams:
\begin{equation}
\log([OIII]/H \beta) < 0.61 / (\log([NII]/H \alpha) - 0.05) +1.3,
\end{equation}
\begin{equation}
\log([OIII]/H \beta) < 0.72 / (\log([SII]/Ha) - 0.32) + 1.30, 
\end{equation}
\noindent
and
\begin{equation}
\log([OIII]/H \beta) < 0.73 / (\log([OI]/Ha) + 0.59) + 1.33
\end{equation}

2. {\bf Composite galaxies} lie between the Ka03 and Ke01 classification lines on the \NIIHa\ vs \OIIIHb\ diagram:
\begin{equation}
 0.61 / (\log([NII]/H \alpha) - 0.05) +1.3 < \log([OIII]/H \beta),
\end{equation}
\begin{equation}
0.61 / (\log([NII]/Ha) - 0.47) + 1.19 > \log([OIII]/H \beta), 
\end{equation}

3. {\bf Seyfert galaxies} lie above the Ke01 classification line on the \NIIHa, \SIIHa, and \OIHa\ diagnostic diagrams and above the Seyfert-LINER line on the \SIIHa\ and \OIHa\ diagrams, i.e.
\begin{equation}
0.61 / (\log([NII]/Ha) - 0.47) + 1.19 < \log([OIII]/H \beta), 
\end{equation}
\begin{equation}
0.72 / (\log([SII]/Ha) - 0.32) + 1.30 < \log([OIII]/H \beta), 
\end{equation}
\begin{eqnarray}
 0.73 / (\log([OI]/Ha) + 0.59) + 1.33 < \log([OIII]/H \beta)\\
(OR\,\, \log([OI]/Ha) > -0.59)
\end{eqnarray}
\noindent
and
\begin{equation}
1.89 \log([SII]/Ha) + 0.76 < \log([OIII]/H \beta), 
\end{equation}
\begin{equation}
1.18 \log([OI]/Ha) + 1.30 < \log([OIII]/H \beta).
\end{equation}

4. {\bf LINERs} lie above the  Ke01 classification line on the \NIIHa, \SIIHa, and \OIHa\ diagnostic diagrams and below the Seyfert-LINERs line on the  \SIIHa\ and \OIHa\ diagrams, i.e.

\begin{equation}
0.61 / (\log([NII]/Ha) - 0.47) + 1.19 < \log([OIII]/H \beta) 
\end{equation}
\begin{equation}
0.72 / (\log([SII]/Ha) - 0.32) + 1.30 < \log([OIII]/H \beta), 
\end{equation}
\begin{equation}
 \log([OIII]/H \beta) < 1.89 \log([SII]/Ha) + 0.76, 
\end{equation}
\begin{eqnarray}
0.73 / (\log([OI]/Ha) + 0.59) + 1.33 & < \log([OIII]/H \beta)\\
 (OR \,\,  \log([OI]/Ha) > -0.59)
\end{eqnarray}
\begin{equation}
 \log([OIII]/H \beta)< 1.18 \log([OI]/Ha) + 1.30, 
\end{equation}

5. {\bf Ambiguous galaxies} are those that are classified as one type of object in one or or two  diagrams and classified as another type of object in the remaining diagram(s).  In our scheme, ambiguous galaxies fall into one of two categories: (a) galaxies that lie in the Seyfert region in either the \SIIHa\ or \OIHa\  diagram and in the LINER region in the remaining (\OIHa\ or \SIIHa) diagram, or (b) galaxies that lie in the composite region (below the Ke01 line) in the \NIIHa\ diagram but that lie above the Ke01 line in either the \SIIHa\ or \OIHa\ diagram.

 According to this scheme, our  85224-galaxy sample contains 63893 (75\%) star-forming galaxies, 
2411 (3\%) Seyferts, 6005 (7\%) LINERs, and 5870 (7\%) composites.  The remaining galaxies are 
ambiguous galaxies (7045; 8\%).

\subsubsection{Simple Diagnostic Diagram}

\begin{figure*}
\includegraphics[width=8.5cm]{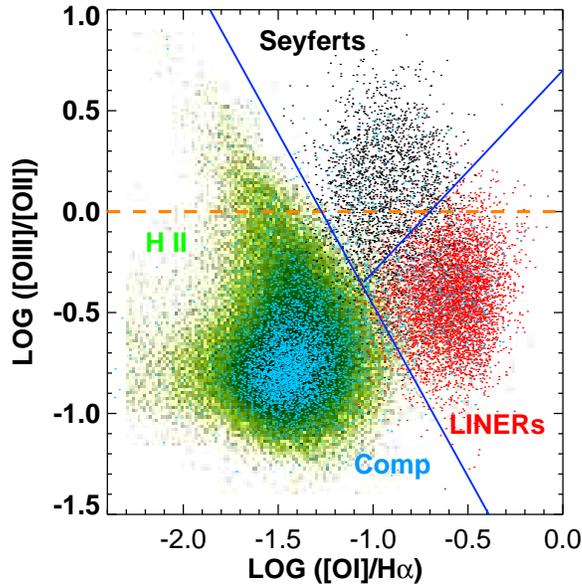}
\caption{The \OIIIOII\ vs \OIHa\ diagnostic diagram for SDSS galaxies with 
$S/N > 3$.   Galaxies have been classified using the standard BPT diagnostic diagrams (Figure~\ref{class_diag}).  Ambiguous galaxies are not included.  Our new preferred classification 
scheme is shown in blue.  The \citet{Heckman80} LINER line (purple dashed) is also shown.
\label{OIIIOII_OIHa}}
\end{figure*}

Using the classifications obtained in the previous section, we investigate other line diagnostic diagrams that may be able to separate the different classes in a simpler way.  Figure~\ref{OIIIOII_OIHa} shows the \OIIIOII\ vs \OIHa\ diagnostic diagram for the \mbox{H\,{\sc ii}}-region like 
Seyferts, LINERs, and composites.  We exclude ambiguous objects from this plot.  LINERs 
and Seyferts form two distinct groups on this diagram and both groups are easily separated from 
the \mbox{H\,{\sc ii}} region-like galaxies and composites.  The reason for this clean separation is twofold; \OIIIOII\ 
is a sensitive diagnostic of the ionization parameter of the gas, while \OIHa\ is sensitive to the hardness of the ionizing radiation field.  The ionization parameter is a measure of the amount of ionization that 
a radiation field can drive as it moves through the nebula.  Seyfert galaxies have a higher ionization 
parameter than LINERs (by definition) or star-forming galaxies and therefore Seyferts separate vertically (\OIIIOII) from the other classes of objects.   Both Seyfert galaxies and LINERs have hard power-law ionizing radiation fields and thus separate from star-forming galaxies in the horizontal 
(\OIHa) direction.  Interestingly, composite galaxies lie within the star-forming sequence on this diagram, indicating that the hardness of the ionizing radiation field, and the ionization parameter of composites are likely to be dominated by their star-formation.

To calculate an empirical separation between LINERs,
Seyferts, and \mbox{H\,{\sc ii}}$+$composites, we find the minima of a 2-dimensional histogram of 
\OIIIOII\ and \OIHa.  These minima are fit with a least-squares line of best-fit.  The resulting empirical separations are shown in blue.  For comparison, the \citet{Heckman80} classification line is shown (purple dashed line).  Clearly the Heckman classification scheme would separate most LINERs from Seyfert galaxies.   This diagnostic diagram is a more simple method for separating LINERs, Seyferts, and star-forming galaxies (including composites).  Our separations are given by;\\

{\bf \mbox{H\,{\sc ii}}\, \&\, Composites:}
\begin{equation}
\log ([OIII]/[OII] <   -1.701 \log([OI]/H\alpha)-2.163
\end{equation}
{\bf LINERs:}
\begin{equation}
  -1.701 \log([OI]/H\alpha)-2.163 <  \log ([OIII]/[OII])
\end{equation}
\begin{equation}
 \log ([OIII]/[OII] <  1.0 \log([OI]/H\alpha) + 0.7
\end{equation}
{\bf Seyferts:}
\begin{eqnarray}
   -1.701 \log([OI]/H\alpha)-2.163 < & \log ([OIII]/[OII])\\
 1.0 \log([OI]/H\alpha) + 0.7< & \log ([OIII]/[OII])
\end{eqnarray}

Note that the \OIIIOII\ vs \OIHa\ diagram should not be used to separate \mbox{H\,{\sc ii}} galaxies from composite objects and it relies on accurate reddening correction between [\mbox{O\,{\sc iii}}]~$\lambda 5007$ and 
\OII~$\lambda \lambda 3727,29$.   

\begin{figure*}
\begin{minipage}{155mm}
\includegraphics[width=15cm]{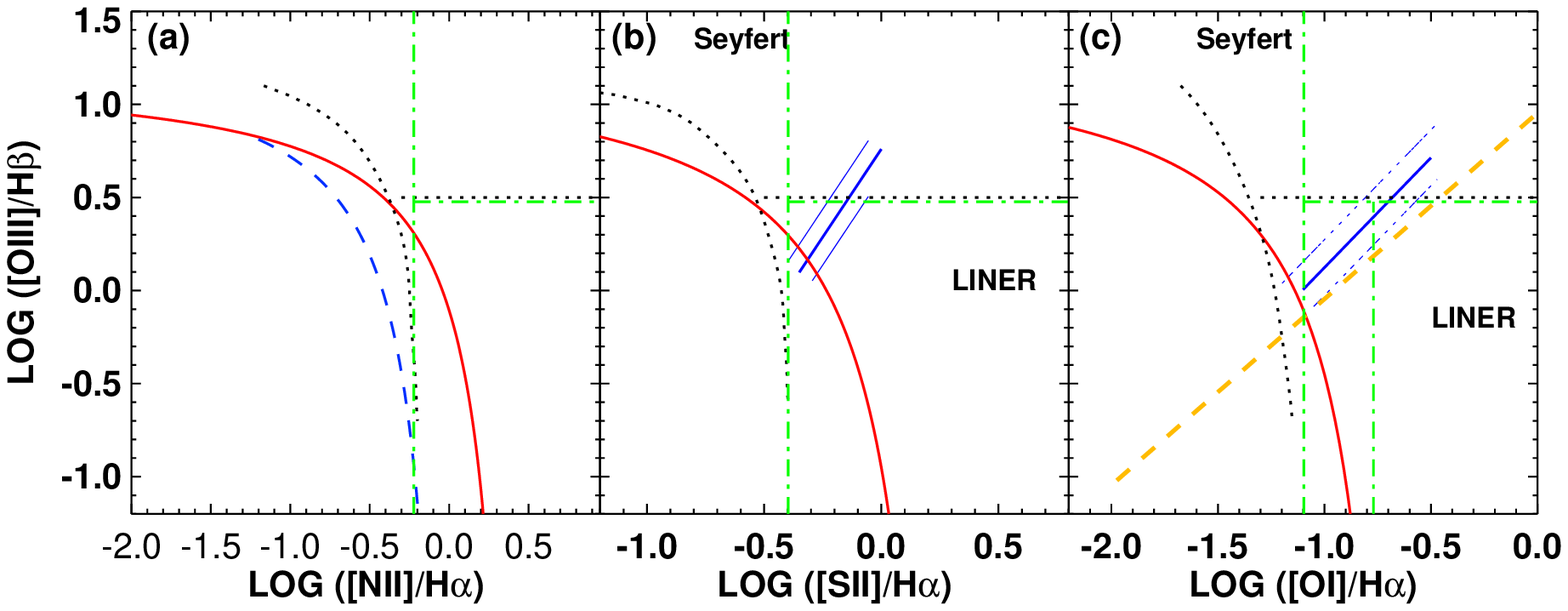}
\caption{The three BPT diagrams showing different methods for classifying galaxies using 
emission-line ratios.  The \citet{Kewley01a} extreme starburst classification line (red solid), 
the \citet{Kauffmann03b} pure star-formation line (blue dashed), our new 
Seyfert-LINER line (blue solid), the \citet{Veilleux87} classification scheme (black dotted), 
the \citet{Heckman80} LINER line (purple dashed), and the \citet{Ho97} classification schemes 
(green dot-dashed) are shown.
\label{class_comp}}
\end{minipage}
\end{figure*}

\subsubsection{Comparison with Previous Diagnostic Schemes}

In Figure~\ref{class_comp} we compare our new classification scheme with previous methods of classification.   The solid green line shows where the \citet[][; hereafter H80]{Heckman80} LINER 
classification line lies in relation to our Seyfert-LINER line.  H80 defined LINERs as having \OII~$\lambda 3727 \ga$~[\mbox{O\,{\sc iii}}]~$\lambda 5007$ and \OI~$\lambda 6300 \ga 1/3$~[OIII]~$\lambda5007$.    The H80 line has a very similar slope to our Seyfert-LINER line and 
is close to the 0.1 dex error markers for our line.  The H80 LINER criteria are sometimes used in 
combination with additional line ratio criteria using \OIIIHb\ and/or \NIIHa, \SIIHa, \OIHa\ to remove star-forming galaxies (e.g., \citealt{Heckman83}).   
Using our "extreme starburst line" to remove galaxies 
dominated by their star formation (including composites), we find that all galaxies that are classed as LINERs using the H80 criteria are also classed as LINERs using our scheme.  Conversely, approximately 56\% of our LINERs are also classified as LINERs using the H80 scheme.  
The remaining galaxies lie above the H80 line but below our Seyfert-LINER dividing line.  If we do not
remove star-forming galaxies and we impose the H80 criteria (\OII~$\lambda 3727 \ga$~[\mbox{O\,{\sc iii}}]~$\lambda 5007$ and \OI~$\lambda 6300 \ga 1/3$~[OIII]~$\lambda5007$), then 
15\% of galaxies that are classed as LINERs using the H80 criteria are also classed as LINERs in our scheme.  The remaining 85\% are \mbox{H\,{\sc ii}} galaxies (68\%), composites (7\%), and ambiguous galaxies 
(9\%).  Alternatively, $\sim$78\% of our LINERs are also classed as LINERs using the H80 scheme.

An alternative classification scheme was proposed by VO87 for galaxies where \OII\ is not 
measured.  The VO87 scheme is shown in Figure~\ref{class_comp} (black dotted lines).  If our 
classification is correct, galaxies classified as LINERs according to the VO87 scheme are
either true LINERs, Seyferts, or composites.   Of the galaxies classified as LINERs
in the VO87 scheme, 67\% are LINERs, 5\% are Seyferts, 4\% are composites, 0.01\% are \mbox{H\,{\sc ii}} 
region-like, and 23\% are 
ambiguous objects.  These ambiguous objects are likely to be composites or transition objects because they lie within the composite region in one or two of the BPT diagrams and in the AGN region in the remaining diagrams.   On the other hand, approximately 87\% of our LINERs are also classified as LINERs using the VO87 method.

\citep[][; hereafter HFS97]{Ho97} defined a new classification scheme for the four different classes of objects using the nuclear emission-line ratios of 418 galaxies.   Their classification scheme is shown on Figure~\ref{class_comp} as green dot-dashed lines.  The HFS97 LINERs criterion for \OIHa\  and 
\OIIIHb\ begins on our Seyfert-LINER classification line.  Because of this LINER criterion, 
92\% of galaxies classed as LINERs in the HFS97 scheme are also LINERs in our scheme.  The 
remaining 8\% are ambiguous objects that lie within our LINER region on one or more diagrams 
and within the Seyfert region on the remaining diagrams. 

To conclude, because $\sim$90-100\% of galaxies classified using H80 (with reliable removal of \mbox{H\,{\sc ii}} galaxies) or HFS97 as LINERs remain 
LINERs in our scheme, previous studies of LINERs defined according to H80 or HFS97 
should reflect the true properties of LINERs.   In contrast, $\sim 1/3$ of objects in LINER  studies that 
have used the VO87 scheme may be AGN-\mbox{H\,{\sc ii}} composites and Seyferts.  Without pre-removal of  \mbox{H\,{\sc ii}} galaxies, the H80 scheme includes a substantial fraction (68\%) of \mbox{H\,{\sc ii}} galaxies.  The relative 
ratio of composite, Seyfert, and true LINERs will depend on any additional selection criteria used in previous LINERs surveys (such as luminosity or color selection).

\begin{figure}
\includegraphics[width=8.5cm]{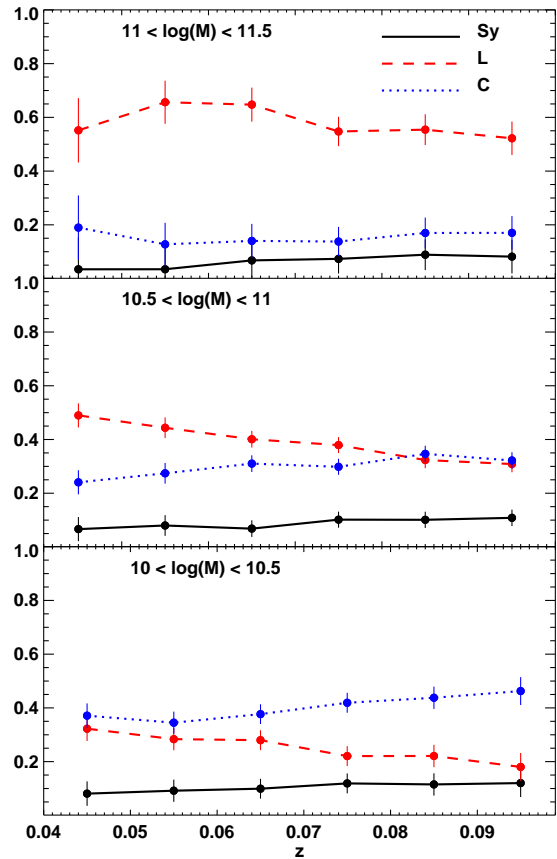}
\caption{The fraction of galaxies classified as Seyfert (black solid line), LINER (red dashed line), 
and composite (blue dotted line) as a function of redshift within three mass ranges.  Ambiguous 
galaxies are not included in this plot.
\label{classfrac_z}}
\end{figure}

\subsubsection{Completeness and Aperture effects as a Function of Class}\label{Ap_class}

At redshifts $z< 0.01$ we can detect galaxies with ${\rm M}_* > 10^{10} {\rm M}_{\odot}$ irrespective of their stellar
mass-to-light ratios.  Thus we only consider galaxies with masses greater than $10^{10} {\rm M}_{\odot}$ in this analysis.   We note that the fraction of galaxies with masses less than $10^{10} {\rm M}_{\odot}$  is small (4\%).
     
In Figure~\ref{classfrac_z}, we show how the fraction of non-\mbox{H\,{\sc ii}} galaxies classed as Seyfert, LINER and composite changes over the redshift range of our sample.  Ambiguous galaxies are not shown.
The fraction of LINERs falls with redshift from $z\sim 0.045$ to $z\sim 0.1$ in the $11<{\rm log(M)}<11.5$ mass range, $z\sim0.03$ to $z\sim 0.1$ in the 
$10.5<{\rm log(M)}<11$ mass range, and $z\sim0.1$ to $z\sim 0.1$ in the $10<{\rm log(M)}<10.5$ mass range.  This drop in LINER fraction is similar to the slope that \citet{Kauffmann03b} found for low luminosity AGN (log(L[OIII])$<7$~\Lsun).  The drop occurs because weak emission lines become increasingly difficult to detect as the physical aperture subtended by the SDSS fibre becomes larger.
Over the redshift range $0.04 < z < 0.1$, the change in LINER fraction is not very large, so aperture bias should not affect our conclusions substantially.

\begin{figure}
\includegraphics[width=8.5cm]{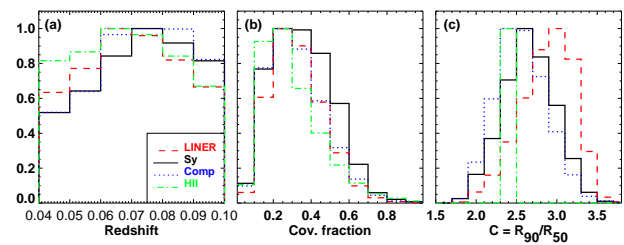}
\caption{The distribution of (a) redshift, (b) g-band fiber covering fraction, and (c) concentration for non-\mbox{H\,{\sc ii}} galaxies classified as Seyfert (black solid line), LINER (red dashed line), and composite (blue dotted line) as a function of redshift within for mass ranges.  Ambiguous galaxies are not included in this plot.
\label{comp_effects}}
\end{figure}

Figure~\ref{comp_effects} shows the overall redshift distribution, g-band fiber covering fraction, 
and concentration index of each spectral type.  

 LINERs have redshift distributions skewed towards lower redshifts than Seyferts or composites, as already seen in Figure~\ref{classfrac_z}.  Figure~\ref{comp_effects}c indicates that the LINERs in our sample are more concentrated than Seyferts or composites.  LINERs have only a slightly lower mean g-band fiber covering fraction for LINERs ($0.34\pm 0.01$) than for Seyfert galaxies ($0.36 \pm 0.01$) Figure~\ref{comp_effects}b).    Because the difference in g-band covering fraction between Seyferts and LINERs is small, we proceed with our analysis of the host properties of LINERs and Seyferts.

\section{Star-forming Distance}\label{dist}

\begin{figure}
\includegraphics[width=8.5cm]{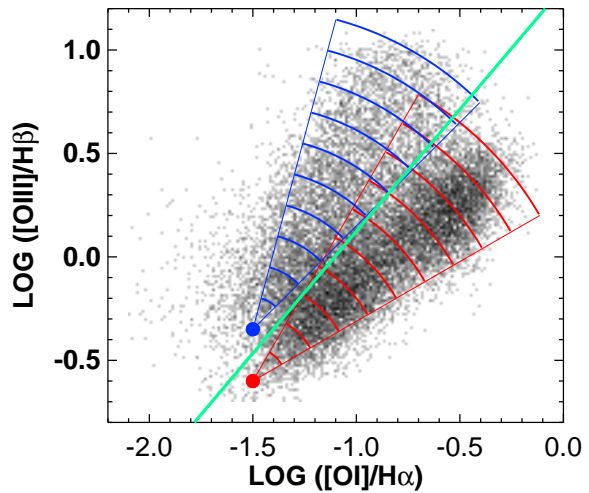}
\caption{The \OIIIHb\ vs \OIHa\ diagnostic diagram for the AGN in the SDSS sample.  Colored 
curves correspond to our empirical definition of distance from the star-forming sequence \SFD.  Blue and red curves give lines of constant \SFD\ for Seyferts and LINERs respectively.  The green line shows out dividing line between Seyferts and LINERs.
\label{SFdist}}
\end{figure}

To explore the host properties as a function of distance from the star-forming sequence, we define an 
empirical linear distance (in dex) from the star-forming sequence for both the Seyfert and LINER branches on the \OIIIHb\ vs \OIHa\ diagnostic diagram shown in Figure~\ref{SFdist}
A base point (red and blue circles in Figure~\ref{SFdist}) is defined for each branch.  A peak radius
of 1.55~dex from the base point is defined for Seyferts.  The peak radius is 1.65~dex for LINERs.  The distance between the peak radius and the base point for each branch is divided into ten bins.  The position of the base and peak points have been chosen to ensure that the first and last bins contain atleast 50 data points.

Note that because \SFD\  is defined in dex in log line-ratio space, it does not give the fraction of star-formation or AGN emission in a galaxy.  

\begin{figure}
\includegraphics[width=8.5cm]{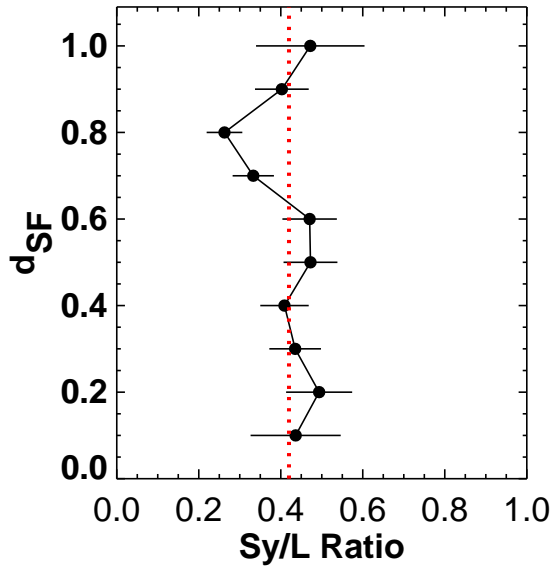}
\caption{The ratio of Seyferts to LINERs as a function of \SFD.  The dotted line shows the mean Seyfert to LINER ratio. 
\label{SyL}}
\end{figure}

In Figure~\ref{SyL} we show the relative ratio of Seyferts and LINERs as a function of \SFD. In this plot, we 
include composite Seyfert+\mbox{H\,{\sc ii}} and LINER+\mbox{H\,{\sc ii}} (classified using our Seyfert-LINER dividing line) as Seyferts and LINERs respectively.   Only pure star-forming galaxies classified according to the Ka03 
classification line (equation 1) and LINER/Seyfert ambiguous galaxies are excluded from Figure~\ref{SyL}.  The composite-pure AGN boundary corresponds roughly to \DSF$=0.3-0.4$.    

Figure~\ref{SyL} shows that the ratio 
of Seyfert to LINERs remains roughly constant with distance from the star-forming sequence.  This result is 
unexpected because both Seyfert-\mbox{H\,{\sc ii}} and LINER-composites exist and there is a large luminosity 
difference between Seyferts and LINERs.   One might expect that in Seyferts-\mbox{H\,{\sc ii}} composites, the stellar emission would be overwhelmed by the luminous Seyfert emission while in LINERs, stellar emission could dominate in many LINER-\mbox{H\,{\sc ii}} composites.   If this expectation were correct, there should be many more LINER-\mbox{H\,{\sc ii}} composites than Seyfert-\mbox{H\,{\sc ii}} composites.   The fact that the Seyfert to LINER ratio is roughly independent of $d_{SF}$ (Figure~\ref{SyL}) implies that star formation and AGN are coupled such that more powerful AGN also have stronger star formation. 

\section{Host Properties }\label{host}

Using our new classification scheme, we explore the host properties of the AGN in our sample.  
The host properties of AGN in the SDSS were previously studied by \citet{Kauffmann03b}.  They 
used the [OIII] luminosity to discriminate between Seyferts and low-luminosity AGN.   The following
analysis extends this work by comparing the host properties of LINERs, Seyferts and composites as a function of distance from the star-forming sequence.   

\subsection{Stellar Population Age }\label{age}

We use the 4000\AA\ break ($D4000$) and the H$\delta$ Balmer absorption line as indicators of the 
age of the stellar population.   The 4000\AA\ break is created by absorption lines located around 
4000\AA\ and was first defined by \citet{Bruzual83}, and recently redefined by \citet{Balogh99} as:

\begin{equation}
D4000 = \frac{\int^{4100}_{4000} f_{\lambda} d_{\lambda}}{\int^{3950}_{3850} f_{\lambda} d{\lambda}}
\end{equation}

The strength of the 4000\AA\ break is influenced by temperature and metallicity.    As the temperature 
of the stellar atmospheres decreases with age, the metal opacity strengthens and \D4\ becomes large.
D4000 was calibrated empirically by \citet{Gorgas99} and theoretically by \citet{Bruzual03}.  
The theoretical calibrations predict that D4000 provides a reliable estimate of the galaxy age for galaxies with mean stellar ages less than a few Gyr.  In the presence of current or recent star formation, D4000 is small because the metal opacity in the atmospheres of O and B stars weakens.

A complementary age indicator is the equivalent width (EW) of the \Hd\ absorption line. The \Hd\ absorption line is produced in the atmospheres of A to G stars and is useful for predicting the age of bursts that ended 4Myr - 1 Gyr ago \citep{Gonzalez99,Worthey97}.   We use the \HdA\ index defined by \citet{Worthey97}:

\begin{equation}
{\rm H}\delta_{A} = ( 4091.00  -  4112.25)/( 1 - F_{I}/F_{C})
\end{equation}

where $F_{I}$ is the flux within the (4091.00 - 4112.25) bandpass  and $F_{C}$ is the pseudo-continuum 
flux within the bandpass.   

\begin{figure}
\includegraphics[width=8.5cm]{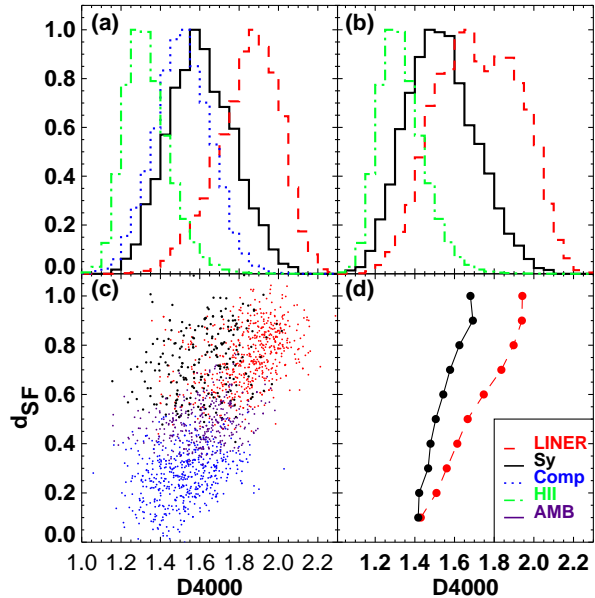}
\caption{(a) The distribution of D4000 for LINERs (red dashed), Seyferts (black solid),
composites (blue dotted) and \mbox{H\,{\sc ii}} region-like galaxies (green dot-dashed). (b) The distribution 
of D4000 for LINERs (red-dashed), Seyferts (black solid), and \mbox{H\,{\sc ii}} region-like galaxies (green 
dot-dashed) where LINERs and Seyferts include LINER+\mbox{H\,{\sc ii}} and Seyfert+\mbox{H\,{\sc ii}} composites.  
Only pure star-forming galaxies classified according to the Ka03 
classification line (equation 1) and LINER/Seyfert ambiguous galaxies are excluded.
(c) D4000 for a uniform random sampling of the Seyferts (black) and LINERs (red), composites (blue) and ambiguous galaxies (purple) 
as a function of distance from the Ke01 line.  (d) The median D4000 for Seyferts (black) and LINERs (red) as a function of distance from the Ke01 line, including Seyfert+\mbox{H\,{\sc ii}} and LINER+\mbox{H\,{\sc ii}} galaxies.  A distance of 1.0 indicates that the optical line ratios are dominated by  ionizing radiation from the Seyfert or LINER nucleus.  
\label{D4000_SFdist}}
\end{figure}

\begin{figure}
\includegraphics[width=8.5cm]{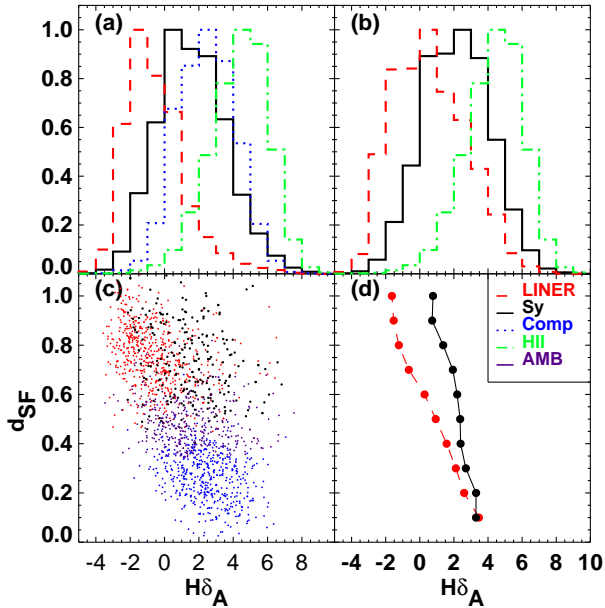}
\caption{(a) The distribution of \HdA\ for LINERs (red dashed), Seyferts (black solid),
composites (blue dotted) and \mbox{H\,{\sc ii}} region-like galaxies (green dot-dashed). (b) The distribution 
of \HdA\ for LINERs (red-dashed), Seyferts (black solid), and \mbox{H\,{\sc ii}} region-like galaxies (green 
dot-dashed) where LINERs and Seyferts include LINER+\mbox{H\,{\sc ii}} and Seyfert+\mbox{H\,{\sc ii}} composites.
(c) \HdA\ for a uniform random sampling of the Seyferts (black) and LINERs (red), composites (blue) and ambiguous galaxies (purple) 
as a function of distance from the Ke01 line.  (d) The median \HdA\ for Seyferts (black) and LINERs (red) as a function of distance from the Ke01 line, including Seyfert+\mbox{H\,{\sc ii}} and LINER+\mbox{H\,{\sc ii}} galaxies.  A distance of 1.0 indicates that the optical line ratios are dominated by  ionizing radiation from the Seyfert or LINER nucleus.  
\label{Hdelta_SFdist}}
\end{figure}

The evolution of D4000 and \HdA\ have been investigated by \citet{Kauffmann03d}  using high resolution spectral libraries and the \citet{Bruzual03} stellar population synthesis models.  They find 
that neither index depends strongly on metallicity until at least 1~Gyr after the burst.

In Figure~\ref{D4000_SFdist}a, we show the normalized distribution of D4000 for LINERs (red).  For comparison, we plot the D4000 distributions for Seyferts (black), composites (blue), and \mbox{H\,{\sc ii}} region-like galaxies (green).  The stellar population of LINERs is older than the other galaxy types.  Seyferts and composites have a similar range of 
stellar ages, and \mbox{H\,{\sc ii}} region-like galaxies are dominated by relatively young star-forming regions, as 
expected. 

To avoid biasing our analysis against galaxies that contain significant star formation,  in Figure~\ref{D4000_SFdist}b we give the normalized distribution of D4000 without defining composites as a separate class.  Rather, we use our Seyfert-LINER classification line to divide the composite galaxies into \mbox{H\,{\sc ii}}+LINER and \mbox{H\,{\sc ii}}+Seyfert classes.   We include \mbox{H\,{\sc ii}}+LINER and \mbox{H\,{\sc ii}}+Seyfert in the LINER and Seyfert classes respectively.     Only pure star-forming galaxies classified according to the Ka03 
classification line (equation 1) and LINER/Seyfert ambiguous galaxies are excluded.
Comparison of panels (a) and (b) shows that while
  the difference between Seyferts and LINERs is somewhat reduced
  if the composites are included, there is still a very clear age offset between the two classes.
  
 In Figure~\ref{D4000_SFdist}c, we show how the stellar population evolves as a function of \SFD\ for Seyferts, composites, LINERs, and ambiguous galaxies.   Figure~\ref{D4000_SFdist}d gives the median D4000 at each 0.1 interval of \SFD.  Composites are included in the LINER and Seyfert classes as in Figure~\ref{D4000_SFdist}b.  Pure LINERs and Seyferts correspond roughly to a \SFD\ of 0.5.   The correlation coefficients and 
corresponding significance of the correlation are given in Table~\ref{corr}.  Clearly \SFD\ and D4000 are 
strongly correlated for both LINERs and Seyferts.  Both types of objects have young stellar populations
typical of active star-forming galaxies at low \SFD, by definition.  Both Seyfert and LINER hosts have older stellar populations at large \SFD.  LINERs hosts have an older stellar population than Seyferts at all but the smallest \SFD.  Figure~\ref{Hdelta_SFdist} shows that a similar trend holds for \HdA.   

\subsection{Stellar Mass }\label{mass}

\begin{figure}
\includegraphics[width=8.5cm]{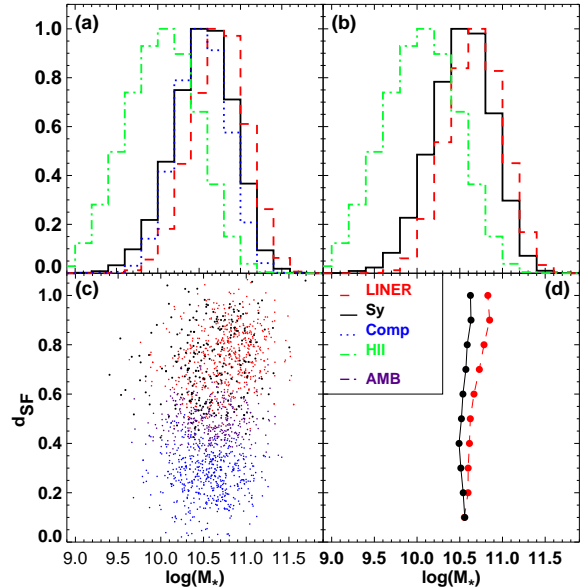}
\caption{(a) The distribution of the logarithm of the stellar mass ($M_{*}$) for LINERs (red dashed), Seyferts (black solid),
composites (blue dotted) and \mbox{H\,{\sc ii}} region-like galaxies (green dot-dashed). (b) The distribution 
of $M_{*}$ for LINERs (red-dashed), Seyferts (black solid), and \mbox{H\,{\sc ii}} region-like galaxies (green 
dot-dashed) where LINERs and Seyferts include LINER+\mbox{H\,{\sc ii}} and Seyfert+\mbox{H\,{\sc ii}} composites.
(c) $M_{*}$ for a uniform random sampling of the Seyferts (black) and LINERs (red), composites (blue) and ambiguous galaxies (purple) 
as a function of distance from the Ke01 line.  (d) The median $M_{*}$ for Seyferts (black) and LINERs (red) as a function of distance from the Ke01 line, including Seyfert+\mbox{H\,{\sc ii}} and LINER+\mbox{H\,{\sc ii}} galaxies.  A distance of 1.0 indicates that the optical line ratios are dominated by  ionizing radiation from the Seyfert or LINER nucleus. 
\label{Mass_SFdist}}
\end{figure}

The SDSS stellar masses were derived by \citet{Kauffmann03a}, and \citet{Gallazzi05} using a combination of $z$-band luminosities and Monte Carlo stellar population synthesis fits to 
D4000 and H$\delta$.  The model fits provide powerful constraints on the star formation history and 
metallicity of each galaxy, thus providing a more reliable indicator of mass than assuming a simple mass-to-light ratio.  \citet{Drory05} compared $\sim 17,000$ SDSS spectroscopically derived masses 
with (a) masses derived from population synthesis fits to the broadband SDSS and 2MASS colors, and (b) masses calculated from SDSS velocity dispersions and effective radii.  They concluded that the three methods for estimating mass agree to within $\sim 0.2$~dex over the $10^{8} - 10^{12}$\Msun\ range.    

The stellar mass distribution of SDSS emission-line galaxies was studied previously by \citet{Kauffmann03b}.  They find a strong dependence between the fraction of galaxies containing AGN and stellar mass.  Figure~\ref{Mass_SFdist}a confirms this result.  The stellar mass distribution of 
star-forming galaxies (green dot-dashed line) is shifted substantially towards lower masses compared 
to the mass distributions of composites, Seyferts, or LINERs.   Figure~\ref{Mass_SFdist}a shows  that the stellar mass  of LINER galaxies is slightly higher than for Seyfert galaxies. Composite objects have a very similar mass distribution to Seyfert galaxies but this may be a coincidence; only a small fraction (14\%) of composites lie above the Seyfert-LINER dividing line on both \NIIHa\ and \SIIHa\ diagrams.  
The majority of composites (67\%) lie on the \mbox{H\,{\sc ii}}-LINER sequence and therefore these galaxies may 
simply have stellar masses intermediate between \mbox{H\,{\sc ii}} region-like galaxies and LINERs.  The median stellar mass of the composite galaxies ($log({\rm M}_{*})=10.54$) is within 0.02~dex of 
the average of the median stellar mass for \mbox{H\,{\sc ii}} region-like and LINER classes (10.52).

Figure~\ref{Mass_SFdist}d shows that the stellar mass is roughly constant as a function of \DSF.  
There is only a small correlation between SF distance and stellar mass for Seyferts and a stronger correlation for LINERs (Table~\ref{corr}). 

\subsection{Mass-to-Light Ratio}\label{M/L}

To investigate how the mass of galaxies at a given observed luminosity changes between the different 
spectral types, we use the median z-band mass-to-light (M/L) ratios.  The M/L ratio of galaxies is influenced by the age of the stellar population and the attenuation by dust.

\begin{figure}
\includegraphics[width=8.5cm]{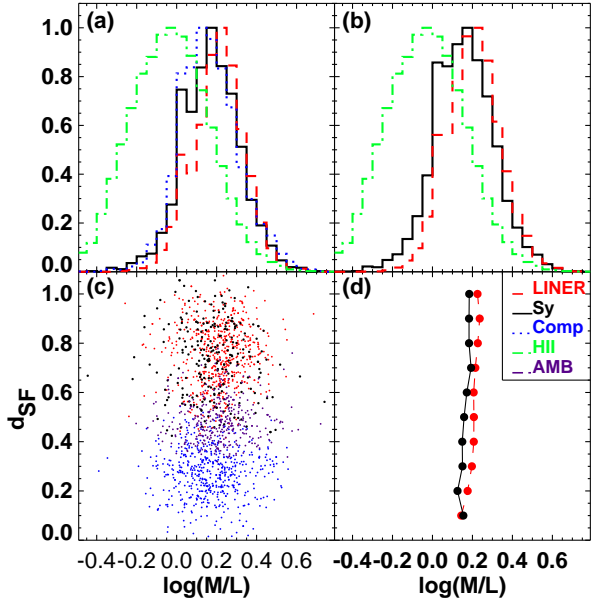}
\caption{(a) The distribution of the logarithm of the stellar mass-to-light ratio (M/L) for LINERs (red dashed), Seyferts (black solid),
composites (blue dotted) and \mbox{H\,{\sc ii}} region-like galaxies (green dot-dashed). (b) The distribution 
of M/L for LINERs (red-dashed), Seyferts (black solid), and \mbox{H\,{\sc ii}} region-like galaxies (green 
dot-dashed) where LINERs and Seyferts include LINER+\mbox{H\,{\sc ii}} and Seyfert+\mbox{H\,{\sc ii}} composites.
(c) M/L for a uniform random sampling of the Seyferts (black) and LINERs (red), composites (blue) and ambiguous galaxies (purple) 
as a function of distance from the Ke01 line.  (d) The median M/L for Seyferts (black) and LINERs (red) as a function of distance from the Ke01 line, including Seyfert+\mbox{H\,{\sc ii}} and LINER+\mbox{H\,{\sc ii}} galaxies.  A distance of 1.0 indicates that the optical line ratios are dominated by  ionizing radiation from the Seyfert or LINER nucleus.
\label{ML_SFdist}}
\end{figure}

The SDSS M/L ratios were derived
by \citet{Kauffmann03a} using Monte Carlo simulations of different star formation histories. 
They find that the M/L ratio correlates strongly with galaxy luminosity such that luminous galaxies have high M/L ratios while faint galaxies have a wide range of M/L ratios.  \citet{Kauffmann03a} used stellar 
evolution models to predict how M/L changes as a function of galaxy luminosity, and stellar population age.  They show that the majority of luminous galaxies  have M/L ratios consistent with having formed stars at  constant rate over a Hubble time, and that faint galaxies have formed most of 
their stars closer to the present.   This effect is reflected in Figure~\ref{ML_SFdist}a where we plot the normalized distribution of M/L for star-forming galaxies, LINERs, Seyferts and Composites.   Seyferts, LINERs and composites have similar M/L distributions, with 
LINERs having only slightly larger M/L ratios on average than Seyferts or composites.  The median M/L ratio is 0.21~dex for LINERs, 0.18~dex for Seyferts and 0.16~dex for composites (the standard error of the median is small; 0.002-0.004~dex).   The M/L range of \mbox{H\,{\sc ii}} region-like galaxies is substantially lower 
than for the other galaxy types, including composites.  The M/L range of \mbox{H\,{\sc ii}} region-like galaxies is 
consistent with their having had present star formation and substantial dust as expected.   In \mbox{H\,{\sc ii}} region-like galaxies, the M/L ratio is correlated with metallicity.  The relatively high M/L ratio of composites may therefore result from their metal-rich stellar population.  The M/L ratio does not change as a function of \SFD\ (Figure~\ref{ML_SFdist}d).

\subsection{Stellar Velocity Dispersion}\label{veldisp}

\begin{figure}
\includegraphics[width=8.5cm]{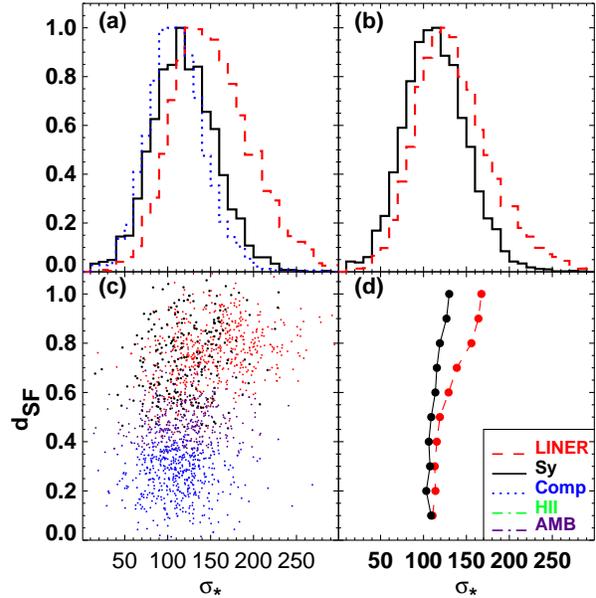}
\caption{(a) The distribution of the logarithm of the stellar velocity dispersion ($\sigma_{*}$) for LINERs (red dashed), Seyferts (black solid),
composites (blue dotted) and \mbox{H\,{\sc ii}} region-like galaxies (green dot-dashed). (b) The distribution 
of $\sigma_{*}$ for LINERs (red-dashed), Seyferts (black solid), and \mbox{H\,{\sc ii}} region-like galaxies (green 
dot-dashed) where LINERs and Seyferts include LINER+\mbox{H\,{\sc ii}} and Seyfert+\mbox{H\,{\sc ii}} composites.
(c) $\sigma_{*}$ for a uniform random sampling of the Seyferts (black) and LINERs (red), composites (blue) and ambiguous galaxies (purple) 
as a function of distance from the Ke01 line.  (d) The median $\sigma_{*}$ for Seyferts (black) and LINERs (red) as a function of distance from the Ke01 line, including Seyfert+\mbox{H\,{\sc ii}} and LINER+\mbox{H\,{\sc ii}} galaxies.  A distance of 1.0 indicates that the optical line ratios are dominated by  ionizing radiation from the Seyfert or LINER nucleus. 
\label{vel_SFdist}}
\end{figure}

The stellar velocity dispersion $\sigma_{*}$ in a galaxy containing an AGN is related to the black hole mass \citep{Tremaine02}.  Figure~\ref{vel_SFdist}a shows the $\sigma_{*}$ distribution for each of the galaxy types and the relationship between $\sigma_{*}$  and \SFD.  LINERs have a broader distribution of \vel\  than Seyferts or composites.  When composite LINER+\mbox{H\,{\sc ii}} and Seyfert+\mbox{H\,{\sc ii}} galaxies are included
in the LINER and Seyfert classes (Figure~\ref{vel_SFdist}b), it is clear that the \vel\ distribution for LINERs has a larger tail towards high \vel\ values.  

Figure~\ref{vel_SFdist}d also shows that \vel\ is roughly constant with  \DSF\ for Seyfert galaxies, with only a slight increase in \vel\ ($\sim 20$~km/s), indicating that black hole mass is relatively constant along the Seyfert branch but that black hole mass may be somewhat larger at high \DSF.  
We show that \vel\ is strongly correlated with the optical line ratios, or \DSF\ for pure LINERs.

\subsection{[OIII] Luminosity}

\begin{figure}
\includegraphics[width=8.5cm]{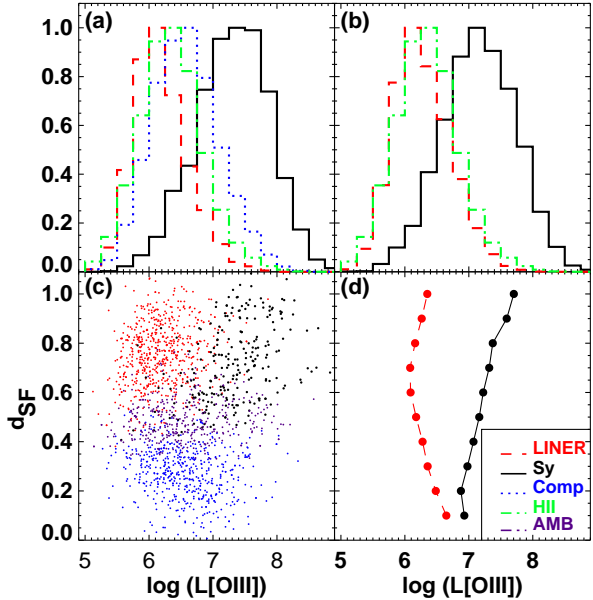}
\caption{(a) The distribution of the logarithm of the extinction-corrected [OIII] luminosity (\Lsun) for LINERs (red dashed), Seyferts (black solid),
composites (blue dotted) and \mbox{H\,{\sc ii}} region-like galaxies (green dot-dashed). (b) The distribution 
of L[OIII] for LINERs (red-dashed), Seyferts (black solid), and \mbox{H\,{\sc ii}} region-like galaxies (green 
dot-dashed) where LINERs and Seyferts include LINER+\mbox{H\,{\sc ii}} and Seyfert+\mbox{H\,{\sc ii}} composites.
(c) L[OIII] for a uniform random sampling of the Seyferts (black) and LINERs (red), composites (blue) and ambiguous galaxies (purple) 
as a function of distance from the Ke01 line.  (d) The median L[OIII] for Seyferts (black) and LINERs (red) as a function of distance from the Ke01 line, including Seyfert+\mbox{H\,{\sc ii}} and LINER+\mbox{H\,{\sc ii}} galaxies.  A distance of 1.0 indicates that the optical line ratios are dominated by  ionizing radiation from the Seyfert or LINER nucleus. 
\label{LOIII_SFdist}}
\end{figure}

In Figure~\ref{LOIII_SFdist}a we show the extinction-corrected [\mbox{O\,{\sc iii}}] luminosity distribution for each spectral class.  LINERs have a lower median [\mbox{O\,{\sc iii}}] luminosity than any of the other spectral 
types; $med(\log {\rm L([OIII])})=6.1$~\Lsun\ for LINERs and $med(\log {\rm L([OIII])})=6.3, 6.5$, and 7.3~\Lsun\ for \mbox{H\,{\sc ii}} region-like, composites and Seyferts respectively.   Seyferts have substantially larger L[OIII]  than 
the other spectral classes from the hard ionizing continuum produced by their AGN.  Composite galaxies have a slightly higher median L[\mbox{O\,{\sc iii}}] than \mbox{H\,{\sc ii}} region-like galaxies and substantially higher median L[OII] than LINERs.  As discussed in Section~\ref{mass}, the majority of composites lie on the LINER branch, but 15\% lie on the Seyfert branch. The large L[OIII] luminosity of these Seyferts raises the median L[OIII] for composites.  

The median [\mbox{O\,{\sc iii}}] luminosity rises as a function of \DSF\ for Seyferts,
The rise in L[\mbox{O\,{\sc iii}}] implies that the ionizing radiation field is 
harder for Seyferts with high \DSF.  \citet{Heckman05} found that the L[OIII] and hard X-ray (3-20 keV) luminosities of AGN are correlated over four orders of magnitude.

In LINERs, the median [\mbox{O\,{\sc iii}}] luminosity falls as a function of \DSF\ until \SFD$\sim0.6$ after which it rises slowly.  The drop in L[\mbox{O\,{\sc iii}}] between \SFD~0-0.6 occurs where \mbox{H\,{\sc ii}}+LINER composites are included in the LINER class.  The star-formation in these \mbox{H\,{\sc ii}}+LINER composites is contributing to the [\mbox{O\,{\sc iii}}] emission until \SFD~0.6.

\subsection{Extinction}\label{extinction}

\begin{figure}
\includegraphics[width=8.5cm]{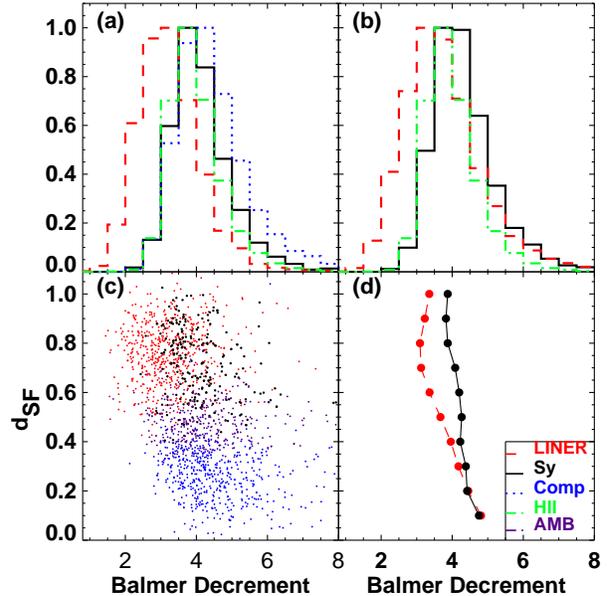}
\caption{(a) The distribution of the logarithm of the Balmer decrement (H$\alpha$/H$\beta$) for LINERs (red dashed), Seyferts (black solid),
composites (blue dotted) and \mbox{H\,{\sc ii}} region-like galaxies (green dot-dashed). (b) The distribution 
of H$\alpha$/H$\beta$ for LINERs (red-dashed), Seyferts (black solid), and \mbox{H\,{\sc ii}} region-like galaxies (green 
dot-dashed) where LINERs and Seyferts include LINER+\mbox{H\,{\sc ii}} and Seyfert+\mbox{H\,{\sc ii}} composites.
(c) H$\alpha$/H$\beta$ for a uniform random sampling of the Seyferts (black) and LINERs (red), composites (blue) and ambiguous galaxies (purple) 
as a function of distance from the Ke01 line.  (d) The median H$\alpha$/H$\beta$ for Seyferts (black) and LINERs (red) as a function of distance from the Ke01 line, including Seyfert+\mbox{H\,{\sc ii}} and LINER+\mbox{H\,{\sc ii}} galaxies.  A distance of 1.0 indicates that the optical line ratios are dominated by  ionizing radiation from the Seyfert or LINER nucleus. 
\label{BD_SFdist}}
\end{figure}

\begin{figure}
\includegraphics[width=8.5cm]{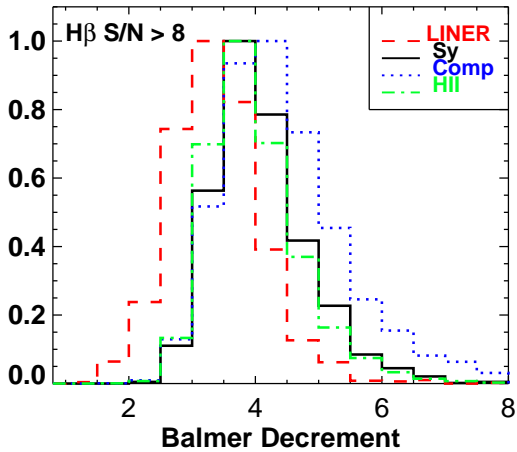}
\caption{The Balmer decrement  distribution for LINERs (red dashed), Seyferts (black solid),
composites (blue dotted) and \mbox{H\,{\sc ii}} region-like galaxies (green dot-dashed) with S/N(\Hb)$> 8$.  
Even LINERs with high S/N have lower Balmer decrements than the other spectral types. 
\label{BD_hist}}
\end{figure}

In Figure~\ref{BD_SFdist} we plot the Balmer decrement for our sample.   The median Balmer decrement of LINERs is significantly smaller than for Seyferts, \mbox{H\,{\sc ii}} region-like galaxies and composites.
A substantial fraction (45\%) of LINERs have Balmer decrements even less than the theoretical value of 3.1, and 33\% of LINERs have Balmer decrements less than the theoretical value for galaxies dominated 
by star formation (2.86).  As discussed in Section~\ref{sample}, a  Balmer decrement
less than the theoretical value can result from a combination of: (1) intrinsically low reddening, (2) 
errors in the stellar absorption correction, and (3) errors in the line flux calibration and 
 measurement.   On the other hand, Balmer decrements lower than 3.1 in AGN can indicate a higher nebular temperature  \citep{Osterbrock89}.   LINER spectra have strong stellar absorption and 
 errors in the absorption correction could bias the Balmer decrement for LINERs, particularly in 
 spectra with low S/N.   We note that many LINERs in our sample have low 
 \Hb\ signal-to-noise (S/N) ratios;  11\% of Seyferts have S/N(\Hb)$< 8$, while 70\% of LINERs have S/N(\Hb)$ < 8$.  This low signal-to-noise may cause the low Balmer decrement in some or perhaps most LINERs but it cannot 
 account for the difference for the entire LINER sample because LINERs with high S/N(\Hb)$> 8$ 
 have a similar distribution of Balmer decrements to those with lower S/N ratios (Figure~\ref{BD_hist}).  \citet{Ho03} found a similar 
 result for the central few hundred parsecs of Seyferts and LINERs and they suggest that Seyfert galaxies contain nuclear regions that are more gas-rich (and dusty) than LINER galaxies.  Our results confirm this conclusion.  

The Balmer decrement is not strongly correlated with distance from the star-forming sequence for LINERs (Figure~\ref{BD_SFdist}b; $r=0.04$), but it is mildly correlated with $d_{SF}$ for Seyferts ($r=-0.17$ with a formally zero probability of obtaining this value by chance) in the sense that Seyferts at large distances from the star-forming seqence contain $\Delta A_V \sim 0.1$~dex less dust on average than 
Seyferts that are close to the star-forming sequence.

\subsection{Host properties across the Seyfert-LINER transition region}

\begin{figure}
\includegraphics[width=8.5cm]{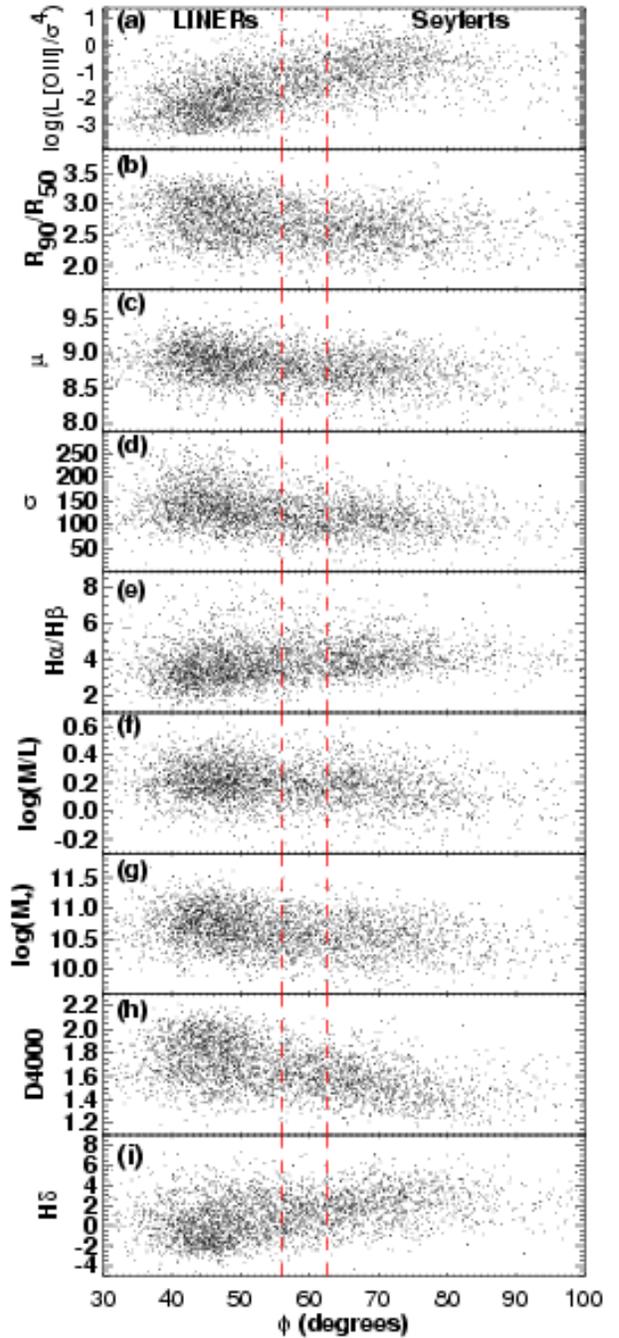}
\caption{The host galaxy properties as a function of angle $\phi$ from the x-axis in the \OIIIHb\ vs \OIHa\ diagnostic diagram.  The Seyfert-LINER boundary is shown as red dashed lines.  Seyferts lie to the right of the boundary while LINERs lie to the left of the boundary.
\label{angle}}
\end{figure}

Given that many of the the host properties of of Seyferts and LINERs are substantially different (D4000, \Hd , \LOIII , etc), it is important to understand whether this difference is a truly bimodal distribution or whether the host properties vary smoothly across the Seyfert-LINER dividing line.  We define $\phi$ as
the angle to the x-axis in the \OIIIHb\ vs \OIHa\ diagram centered on the blue base point in Figure~\ref{arcs}.   Therefore LINERs have small $\phi$ while Seyferts have large $\phi$.   Figure~\ref{angle} shows how the host properties of AGN change across the Seyfert-LINER boundary (red dashed lines).  The host properties clearly do not form a bimodal distribution; they form a smooth 
sequence from Seyfert values to LINER values.  The correlation coefficients and significance for each panel are given in Table~\ref{corr}.  The \LOIIIsig\ (a proxy for the Eddington ratio) correlates most strongly with $\phi$.  We will discuss \LOIIIsig\ as a function of host properties in the following Section.

\section{Host properties as a function of L/L$_{\rm edd}$ }\label{Ledd}


The accretion rate on the black hole is proportional to the bolometric luminosity of the AGN. The stellar velocity dispersion can be used as an estimate of the black hole mass \citep{Tremaine02}.  In type I Seyferts and quasars, the intrinsic [\mbox{O\,{\sc iii}}] luminosity scales with the AGN bolometric luminosity (see \citealt{Heckman04} for a more detailed discussion). It follows that the quantity \LOIIIsig\ is proportional to 
$L/L_{EDD}$.
   
\begin{figure}
\includegraphics[width=8.5cm]{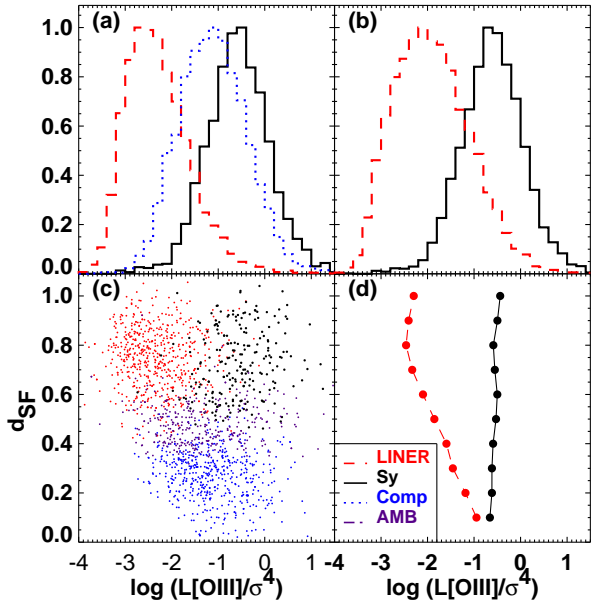}
\caption{(a) The distribution of the logarithm of L[OIII]/$\sigma^4$ (corrected for extinction) for LINERs (red dashed), Seyferts (black solid),
composites (blue dotted) and \mbox{H\,{\sc ii}} region-like galaxies (green dot-dashed). (b) The distribution 
of L[OIII]/$\sigma^4$ for LINERs (red-dashed), Seyferts (black solid), and \mbox{H\,{\sc ii}} region-like galaxies (green 
dot-dashed) where LINERs and Seyferts include LINER+\mbox{H\,{\sc ii}} and Seyfert+\mbox{H\,{\sc ii}} composites.
(c) L[OIII]/$\sigma^4$ for Seyferts (black) and LINERs (red), composites (blue) and ambiguous galaxies (purple) 
as a function of distance from the Ke01 line.  (d) L[OIII]/$\sigma^4$ for Seyferts (black) and LINERs (red) as a function of distance from the Ke01 line, including Seyfert+\mbox{H\,{\sc ii}} and LINER+\mbox{H\,{\sc ii}} galaxies.  A distance of 1.0 indicates that the optical line ratios are dominated by  ionizing radiation from the Seyfert or LINER nucleus. 
\label{LOIIIsig_SFdist}}
\end{figure}

In Figure~\ref{LOIIIsig_SFdist} we show that \LOIIIsig\ does not change as a function of distance 
from the star-forming sequence for Seyfert galaxies.  The \LOIIIsig\ ratio of LINERs is influenced by star-formation (in \LOIII) for \SFD\ between 0-0.6, as in Figure~\ref{LOIII_SFdist}.   Pure Seyferts and LINERs 
(\SFD$\ga 0.6$) have substantially different values of \LOIIIsig.   Figure~\ref{LOIIIsig_SFdist}d indicates that \LOIIIsig\ is roughly constant with \SFD\ for Seyferts.   Both \LOIII\ and $\sigma$ (black hole mass) rise with \SFD for Seyferts, resulting in a constant \LOIIIsig.  The rise in \LOIII\ and $\sigma$ with \SFD\ probably reflects the trend that older, more massive galaxies contain larger black holes.

\begin{figure}
\includegraphics[width=8.5cm]{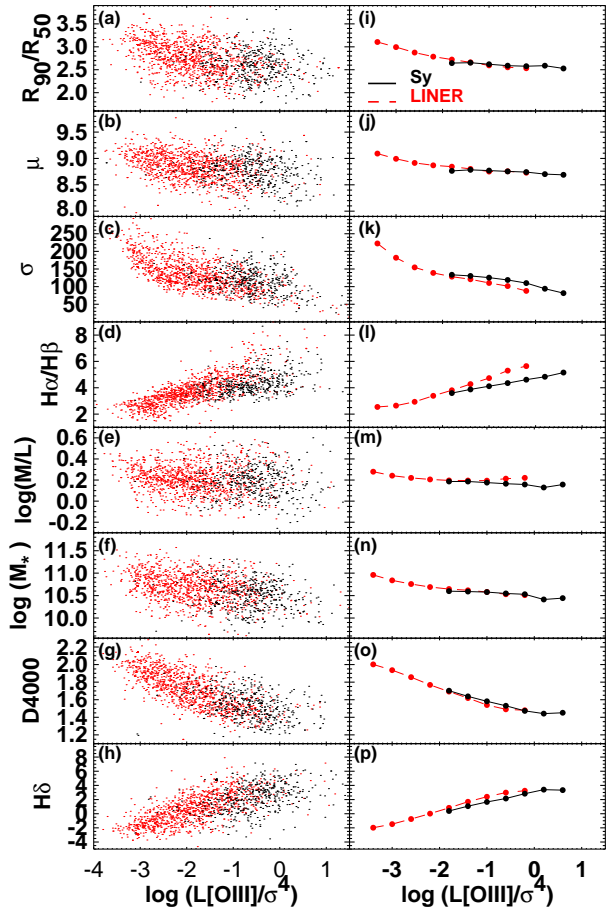}
\caption{\LOIIIsig\ versus (a) concentration, (b) surface mass density, (c) stellar velocity dispersion, (d) Balmer decrement, (e) mass-to-light ratio, (f) stellar mass, (g) the D4000 index, and (h) \Hd\ for LINERs (red), Seyferts (black), and composites (blue).  The right panel shows the median values 
of each host galaxy property for small bins (0.4~dex) of \LOIIIsig.  Only bins containing more than 150 
data points are included.
\label{LOIIIsig}}
\end{figure}

We show how the host galaxy properties of LINERs, Seyferts and composites vary as a function of \LOIIIsig\ in  Figure~\ref{LOIIIsig}.  Remarkably, the relationship between \LOIIIsig\ and all of the host properties for both Seyferts and LINERs forms one smooth sequence.   For \LOIIIsig\ values exhibited by both Seyferts and LINERs ($-1.5<$\LOIIIsig$<-0.5$), the median of the Seyfert and LINER distributions is almost identical for every host property.  This result implies that (1) most (if not all) LINERs in our sample are AGN, and that (2) \LOIIIsig\ is the major fundamental property that divides Seyferts from LINERs and that host property differences are secondary.  For example,  LINERs are found in larger, older galaxies than Seyferts.   Figure~\ref{LOIIIsig} shows that the black hole accretion rates are lower in larger, older galaxies and that mass and age do not need to be physically linked to the differences in the LINER and Seyfert emission line spectra.  

\begin{figure}
\includegraphics[width=8.5cm]{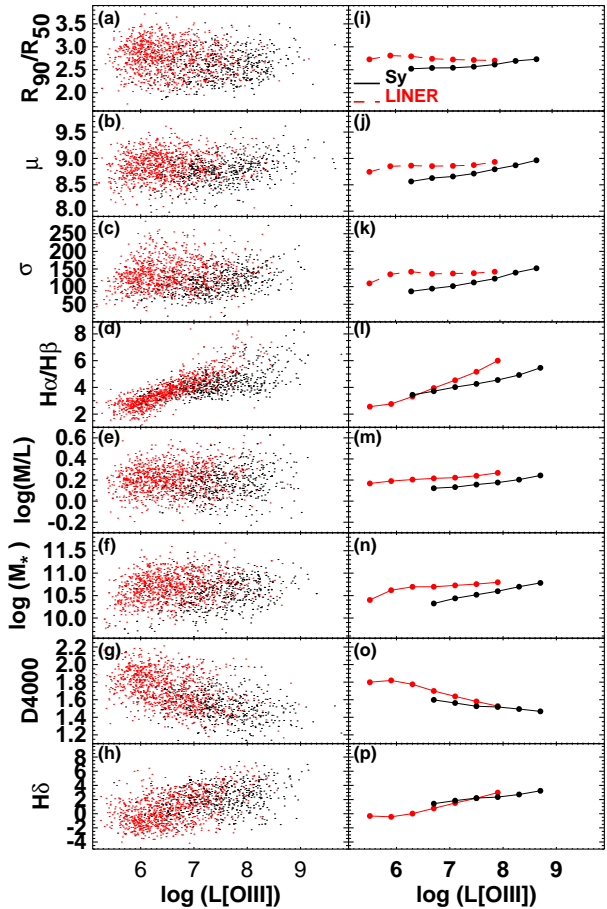}
\caption{\LOIII\ versus (a) concentration, (b) surface mass density, (c) stellar velocity dispersion, (d) Balmer decrement, (e) mass-to-light ratio, (f) stellar mass, (g) the D4000 index, and (h) \Hd\ for LINERs (red), Seyferts (black), and composites (blue).  The right panel shows the median values 
of each host galaxy property for small bins (0.4~dex) of \LOIII.  Only bins containing more than 150 data points are included.
\label{LOIIImulti}}
\end{figure}

We check whether a similar effect occurs as a function of \LOIII\ in Figure~\ref{LOIIImulti}.  All of the relations between host properties and \LOIIIsig\ degrade when \LOIII\ is used in place of \LOIIIsig, confirming that \LOIIIsig is the major property separating Seyferts from LINERs.

\section{Discussion }\label{discussion}

Previous work has shown that that almost all normal galaxies contain black holes and black holes appear to evolve with their host galaxies \citep{Ferrarese00,Gebhardt00}.   In Section~\ref{Ledd} we
showed that although Seyferts and LINERs display different host properties, these differences are largely resolved when the host properties are considered as a function of \LOIIIsig.   Thus the strongest difference between Seyferts and LINERs is \LOIIIsig, or Eddington ratio.   Remarkably, Figure~\ref{LOIIIsig} implies the that if the Eddington ratio of an AGN is known, then its spectral class and broad host galaxy properties can be deduced. 

In Figure~\ref{angle}, we show that the transition in Eddington ratio from Seyferts (high) to LINERs (low) is smooth and that the correlation between classification angle and \LOIIIsig\ is stronger than for any 
of the other host galaxy properties.  This Seyfert-LINER transition may be analogous to the transition between high state and low state black hole accretion in X-ray binary systems (e.g., \citealt{Nowak95} and references therein) as suggested by \citet{Ho05b}.  In X-ray binaries the transition from high to low states is believed to correspond to decreasing mass accretion rates and therefore the primary parameter
in many X-ray binary models is the Eddington ratio (e.g., \citealt{Shakura73,Chen95}).   Theory
suggests that an unstable accretion disk may switch between high and low states \citep{Abramowicz88,Narayan93,Narayan94,Narayan95}.  In this scenario,
the energy released by viscous processes in the high state is radiated and the gas is relatively cool.
 In the low state the viscously released energy may be advected with the gas and radiates inefficiently.  This advection dominated accretion flow (ADAF) is stable to thermal and viscous instabilities \citep{Abramowicz95,Narayan95}.    Because the gas loses little energy through radiation in ADAF models, the gas is extremely hot and the weak spectrum produced is harder than in the high state 
 (e.g., \citealt{Esin97}).   Alternative scenarios to ADAF models can also produce a weak hard spectrum \citep[e.g.,]{Meyer00,Ferreira06}.  Our observed transition between Seyfert and LINERs occurs at \LOIIIsig$\sim 0.10$.  Assuming a bolometric correction of ${\rm L_{BOL}=3500 L[OIII]}_{o}$ \citep{Heckman04}  where \LOIII$_{o}$ is the observed [\mbox{O\,{\sc iii}}] luminosity, and assuming a mean extinction of  ${\rm E(B-V)}=0.3$, our Seyfert-LINER transition corresponds to an Eddington ratio of ${\rm L/L_{EDD}} \sim 0.05$.    This transition ${\rm L/L_{EDD}}$ is within the range of the observed (and theoretical) transition in X-ray binaries; $0.01<{\rm L/L_{EDD}}<0.1$ (e.g., \citealt{Nowak95,Narayan96,Esin97,Barret00,Ferreira06} and references therein).
  
 We investigate the broad, qualitative differences in radiation field between Seyferts and LINERs using the theoretical AGN photoionization models from MAPPINGS III \citep{Groves04a,Groves04b}.   We 
 calculate a simple grid of photoionziation models using four power-law indices ($\alpha = -1.2,-1.4,-1.7,-2.0$) and four ionization parameters ($\log(U)=0,-1.0,-2.0,-3.0$).   The MAPPINGS models are self-consistent and include detailed dust physics and radiation pressure.  We use a metallicity of 2\Zsun and a hydrogen density of 1000~cm$^{-3}$.
 
\begin{figure}
\includegraphics[width=8.5cm]{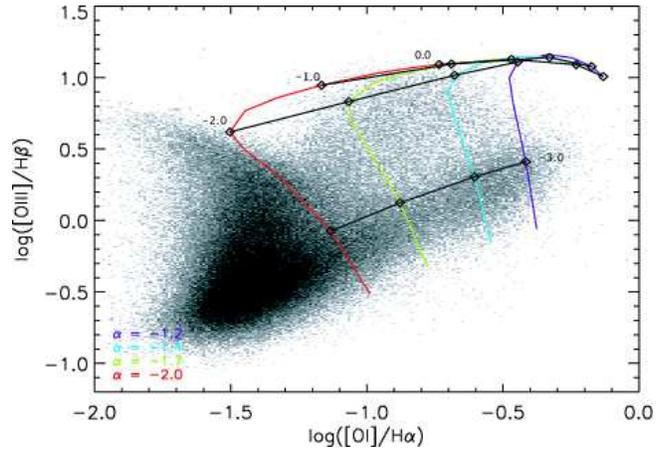}
\caption{The \OIIIHb\ vs \OIHa\ diagnostic diagram showing the SDSS galaxies (black points) and our
theoretical dusty radiation pressure dominated AGN models.  These models were calculated assuming a metallicity of 2\Zsun\ and a hydrogen density of 1000~cm$^{-3}$.  The model grids are given for four 
ionization parameters ($\log(U)=0,-1,-2,-3$) and four power-law indices ($\alpha = -1.2,-1.4,-1.7,-2.0$) as 
marked.  The radiation field in LINERs requires a lower ionization parameter than in Seyfert galaxies and LINERs with strong \OIHa\ require a harder radiation field than Seyferts. \label{Models_OIIIHb_OIHa}}
\end{figure}

\begin{figure}
\includegraphics[width=8.5cm]{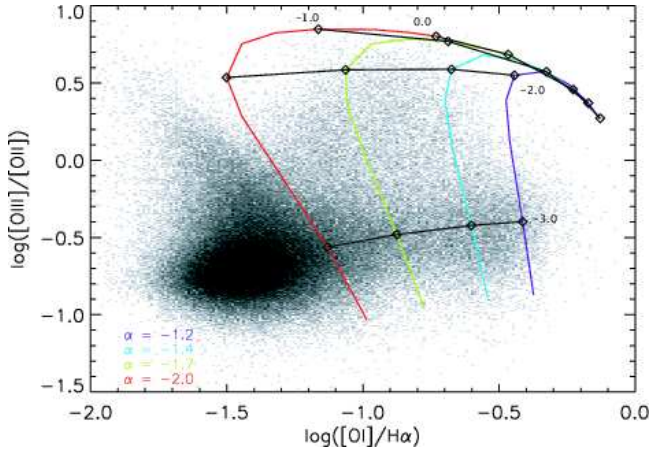}
\caption{The \OIIIOII\ vs \OIHa\ diagnostic diagram showing the SDSS galaxies (black points) and our
theoretical dusty radiation pressure dominated AGN models.  These models were calculated assuming a metallicity of 2\Zsun\ and a hydrogen density of 1000~cm$^{-3}$.  The model grids are given for four 
ionization parameters ($\log(U)=0,-1,-2,-3$) and four power-law indices ($\alpha = -1.2,-1.4,-1.7,-2.0$) as 
marked.  The radiation field in LINERs requires a lower ionization parameter than in Seyfert galaxies and LINERs with strong \OIHa\ require a harder radiation field than Seyferts. \label{Models_OIIIOII_OIHa}}
\end{figure}

 Figure~\ref{Models_OIIIHb_OIHa} shows the position of our AGN relative to the Seyfert and LINER branches on the \OIIIOII\ vs \OIHa\ diagnostic diagram.  The Seyfert branch can be reproduced by an ionizing radiation field with power-law ndex $\alpha= -1.4$ to -1.8 and a high ionization parameter ($\log(U)=-2.5$ to -1.0).  The LINER branch spans a larger range of power-law index, requiring a harder radiation field than Seyferts ($\alpha < -1.4$) at log(\OIHa)$\geq -0.6$.  The LINER branch requires a much lower ionization parameter than the Seyfert branch;  the ionization parameter for LINERs ($\log(U)=-3$) is up to an order of magnitude lower than the ionization parameter typical for Seyferts 
 ($\log(U)=-2$ to -2.5).   Figure~\ref{Models_OIIIOII_OIHa} confirms this result and separates ionization 
 parameter from power-law index over a broader range of ionization parameter.

Figures~\ref{Models_OIIIHb_OIHa} and \ref{Models_OIIIOII_OIHa} indicate that the hardness of the ionizing radiation field increases 
as a function of distance from the star-forming sequence.  We know that Seyferts contain a young-intermediate age stellar population (Figures~\ref{D4000_SFdist}, ~\ref{Hdelta_SFdist}) and that both Seyfert-\mbox{H\,{\sc ii}} and LINER-composites exist.   Given these
results, one would expect that the stellar emission would be overwhelmed by the luminous Seyfert emission in Seyfert-\mbox{H\,{\sc ii}} composites but that stellar emission could dominate in many LINER-\mbox{H\,{\sc ii}} composites.   In this case, there should be many more LINER-\mbox{H\,{\sc ii}} composites than Seyfert-\mbox{H\,{\sc ii}} composites.   The fact that the Seyfert to LINER ratio is independent of $d_{SF}$ (Figure~\ref{SyL}) implies that star formation and AGN are coupled in Seyfert galaxies such that more powerful AGN also have stronger star formation.  Further theoretical and observational work are required to 
confirm this connection between star formation and Seyfert emission.

\section{Conclusions}\label{conclusions}

We have analyzed the emission-line and host properties of 85224 galaxies from the Sloan 
Digital Sky Survey.  We show that Seyferts and LINERs form two separate branches on 
the standard optical diagnostic diagrams.  We present a new optical classification scheme that successfully separates purely star-forming galaxies, Seyferts, LINERs, and composite AGN$+$star-forming galaxies.  We show that  \OIIIOII\ vs \OIHa\  diagnostic diagram easily discriminates between Seyferts, LINERs and galaxies dominated by star-formation (including composites).  We use our new classification scheme to investigate the host properties of AGN galaxies as a function of distance from the star-forming sequence.   We find that:

\begin{itemize}
\item The ratio of Seyferts to LINERs is invariant with distance from the star-forming sequence.  This result implies that star formation and AGN power are coupled.  
\item The stellar populations of Seyferts and LINERs age with distance from the star-forming sequence.  LINERs have a substantially older stellar population than the other spectral types.  The youngest 
stellar population in LINERs has D4000 and \Hd\  values consistent with the oldest stellar population in Seyferts on average.
\item The stellar mass and mass-to-light ratios of Seyferts and LINERs are not strongly correlated with 
distance from the star-forming sequence.  However, LINERs have slightly higher stellar masses and mass-to-light ratios on average than Seyferts or composite objects.
\end{itemize}

We use \LOIIIsig\ as an indicator of the accretion rate in Eddington units.  We compare the host properties of Seyfert and LINER galaxies as a function of \LOIII\ and \LOIIIsig.   We show that:

\begin{itemize}
\item The strongest difference between Seyfert and LINER galaxies is \LOIIIsig.   
\item When the host properties of Seyferts and LINERs are considered in terms of \LOIIIsig, the differences in host properties is resolved and the host properties of Seyferts and LINERs form a smooth sequence with \LOIIIsig.  
\item When the host properties of Seyferts and LINERs are considered in terms of \LOIII, the smooth 
sequence seen with \LOIIIsig\ disappears.
\end{itemize}

These results indicate that most (if not all) LINERs contain AGN, and that the major fundamental difference between Seyferts and LINERs is accretion rate.  We use theoretical AGN photoionization models to show that the LINER branch requires a lower ionization parameter (up to an order of magnitude) than the Seyfert branch, and that strong LINERs (log(\OIHa)$>-1.6$) require a harder ionizing radiation field than galaxies on the Seyfert branch.   These results suggest that the transition between Seyferts and LINERs is analogous to the high-state and low-state transition in X-ray binaries.

\section*{Acknowledgments}
L. J. Kewley is supported by a Hubble Fellowship.  This research has made use of NASA's Astrophysics Data System Bibliographic Services.

 Funding for the creation and distribution of the SDSS Archive has been provided by the Alfred P. Sloan Foundation, the Participating Institutions, the National Aeronautics and Space Administration, the National Science Foundation, the U.S. Department of Energy, the Japanese Monbukagakusho, and the Max Planck Society. The SDSS Web site is http://www.sdss.org/.

The SDSS is managed by the Astrophysical Research Consortium (ARC) for the Participating Institutions. The Participating Institutions are The University of Chicago, Fermilab, the Institute for Advanced Study, the Japan Participation Group, The Johns Hopkins University, Los Alamos National Laboratory, the Max-Planck-Institute for Astronomy (MPIA), the Max-Planck-Institute for Astrophysics (MPA), New Mexico State University, University of Pittsburgh, Princeton University, the United States Naval Observatory, and the University of Washington.


\clearpage
\begin{table}
\caption{Correlation Coefficients  \label{corr}}
\begin{tabular}{lllrr}
\hline
Figure & x-axis & y-axis & corr. coeff\footnotemark[1] & prob\footnotemark[2] (\%) \\
 \ref{SyL} & Sy/L ratio & d$_{\rm SF}$  & -0.36 & 31 \\
\ref{D4000_SFdist}c (LINERs) & D4000 & d$_{\rm SF}$ & 0.74 & 0.0\footnotemark[3] \\
\ref{D4000_SFdist}c (Seyferts) & D4000 & d$_{\rm SF}$ & 0.45 & 0.0\footnotemark[3] \\
\ref{Hdelta_SFdist}c (LINERs) & H$\delta_{\rm A}$ & d$_{\rm SF}$ & -0.67 & 0.0\footnotemark[3] \\
\ref{Hdelta_SFdist}c (Seyferts) & H$\delta_{\rm A}$ & d$_{\rm SF}$ & -0.32 & 0.0\footnotemark[3] \\
\ref{Mass_SFdist}c (LINERs) & log(M$_{*}$)  & d$_{\rm SF}$ & 0.25 & 0.0\footnotemark[3] \\
\ref{Mass_SFdist}c (Seyferts) & log(M$_{*}$)  & d$_{\rm SF}$ & 0.11 & 1$\times 10^{-13}$ \\
\ref{ML_SFdist}c (LINERs) & log(M/L)  & d$_{\rm SF}$ & 0.10 & 0.0\footnotemark[3] \\
\ref{ML_SFdist}c (Seyferts) & log(M/L)  & d$_{\rm SF}$ & 0.12 & 1$\times 10^{-14}$ \\
\ref{vel_SFdist}c (LINERs) & $\sigma_{*}$  & d$_{\rm SF}$ & 0.45 & 0.0\footnotemark[3] \\
\ref{vel_SFdist}c (Seyferts) & $\sigma_{*}$  & d$_{\rm SF}$ & 0.20 & 0.0\footnotemark[3] \\
\ref{LOIII_SFdist}c (LINERs) & log(L([OIII]))  & d$_{\rm SF}$ & -0.13 & 0.0\footnotemark[3] \\
\ref{LOIII_SFdist}c (Seyferts) & log(L([OIII]))  & d$_{\rm SF}$ & 0.38 & 0.0\footnotemark[3] \\
\ref{BD_SFdist}c (LINERs) & H$\alpha$/H$\beta$  & d$_{\rm SF}$ & -0.48 & 24 \\
\ref{BD_SFdist}c (Seyferts) & H$\alpha$/H$\beta$  & d$_{\rm SF}$ & -0.23 & 24 \\
\ref{LOIIIsig_SFdist}c (LINERs) & log(L([OIII])/$\sigma^4$)  & d$_{\rm SF}$ & -0.50 & 0.0\footnotemark[3] \\
\ref{LOIIIsig_SFdist}c (Seyferts) & log(L([OIII])$\sigma^4$)  & d$_{\rm SF}$ & 0.07 & $6\times 10^{-6}$ \\
\ref{angle}a & $\phi$ & log(L([OIII])/$\sigma^4$) & 0.51 &  0.0\footnotemark[3] \\
\ref{angle}b & $\phi$ & R$_{90}$/R$_{50}$ & -0.27 &  0.0\footnotemark[3] \\
\ref{angle}c & $\phi$ & $\mu$ & -0.28 &  0.0\footnotemark[3] \\
\ref{angle}d & $\phi$ & $\sigma$ & -0.33 &  0.0\footnotemark[3] \\
\ref{angle}e & $\phi$ & H$\alpha$/H$\beta$ & 0.28 &  0.0\footnotemark[3] \\
\ref{angle}f & $\phi$ & log(M/L) & -0.21 &  0.0\footnotemark[3] \\
\ref{angle}g & $\phi$ & log(M$_{*}$) & -0.30 &  0.0\footnotemark[3] \\
\ref{angle}h & $\phi$ & D4000 & -0.46 &  0.0\footnotemark[3] \\
\ref{angle}i & $\phi$ & D4000 & 0.37 &  0.0\footnotemark[3] \\
\hline
\end{tabular}
\end{table}

\footnotetext[1]{Spearman-Rank correlation coefficient}
\footnotetext[2]{Probability of obtaining the correlation coefficient value by chance}
\footnotetext[3]{Probability of obtaining the correlation coefficient value by chance is formally zero (1$\times10^{-29}$\%)}

\end{document}